\documentclass[aps,pra,twocolumn,superscriptaddress,notitlepage,nofootinbib,longbibliography]{revtex4-2}
\usepackage{mathrsfs}
\usepackage{epsfig}
\usepackage{graphicx}
\usepackage{amsfonts}
\usepackage[figuresright]{rotating}
\usepackage{amssymb}
\usepackage{amsmath}
\usepackage{dcolumn}
\usepackage{bm}
\usepackage{color}
\usepackage{braket}
\usepackage{units}
\usepackage{xspace}

\usepackage{bbold}
\usepackage[colorlinks=true, allcolors=blue]{hyperref}
\definecolor{mypink3}{cmyk}{0, 0.7808, 0.4429, 0.1412}
\definecolor{mypink1}{rgb}{0.858, 0.188, 0.478}
\definecolor{mypink2}{RGB}{219, 48, 122}

\newcommand{\ZJNU}{Department of Physics, Zhejiang Normal University, Jinhua 321004, People’s Republic of China}
\newcommand{\UC}{Institute for Quantum Science and Technology, University of Calgary, Alberta, Canada T2N 1N4}
\begin{document}
\title{%
Sub-Planck phase-space structure and sensitivity for SU(1,1) compass states}

\author{Naeem Akhtar}
\email{naeem\_abbasi@zjnu.edu.cn}
\affiliation{\ZJNU}
\author{Barry C. Sanders}
\email{sandersb@ucalgary.ca}
\affiliation{\UC}

\author{Gao Xianlong}
\email{ gaoxl@zjnu.edu.cn}
\affiliation{\ZJNU}
\date{\today}
\begin{abstract}
We investigate the sub-Planck-scale structures associated with the SU(1,1) group by establishing that the Planck scale on the hyperbolic plane can be considered as the inverse of the Bargmann index $k$. Our discussion involves SU(1,1) versions of Wigner functions, and quantum-interference effect is easily visualized through plots of these Wigner functions. Specifically, the superpositions of four Perelomov SU(1,1) coherent states (compass state) yields nearly isotropic sub-Planck structures in phase space scaling as $\nicefrac1{k}$ compared with $\nicefrac1{\sqrt{k}}$ scaling for individual SU(1,1) coherent states
and anisotropic quadratically improved scaling for superpositions of two SU(1,1) coherent states (cat state). We show that displacement sensitivity exhibits the same quadratic improvement to scaling.

\end{abstract}

\maketitle

\section{Introduction}\label{sec:introduction}
The quantum uncertainty principle~\cite{Robertson1929,wheeler2014},
arising from commutator relations such as the position-momentum case
$[\hat{x},\hat{p}]=\text{i}\hbar$,
limits the size of a phase-space structure~\cite{wheeler2014},
for example, represented by the Wigner function~\cite{Wig32}
for Heisenberg-Weyl (HW) symmetry~\cite{Wey50}
generated by the HW algebra~$\mathfrak{hw}$(1),
and, more generally, by Moyal symbols for other symmetries~\cite{Moy49}.
The uncertainty principle does not actually mean that the displacement sensitivity,
with ``displacement'' referring to group action on the state,
is limited by this Planck scale
because 
quantum interference in phase space~\cite{Gerry05book,Sch01,drummond2004quantum} yields finer-scale properties. For example, the compass states
(superposition of four coherent states)
have shown sub-Planck spotty structures in Wigner functions over phase space~\cite{Zurek2001}.
Such sub-Planck structures are highly sensitive to environmental decoherence~\cite{Zurek2003} and play a crucial role in the sensitivity of a quantum state against phase-space displacements~\cite{Toscano06,Eff4}.
Sub-Planck structures have been explored in various contexts~\cite{Eff1,Eff2,Eff3,Eff5,Eff6,Eff7,Eff8,Eff9,Eff10,Eff11,Eff12,Eff13,Eff14,Eff15},
and both theoretical~\cite{Prop1,Prop2,Prop3,Prop5,Prop6} and experimental~\cite{Exp1,Exp2,Exp3,Exp4,Exp5} studies of sub-Planck structure have been undertaken.

Coherent states of the harmonic oscillator belong to the dynamical symmetry group, the so-called Heisenberg-Weyl (HW) group~\cite{weyl1950theory}. These coherent states were first introduced by Schr\"{o}dinger in 1926~\cite{Sch26,MN09}
and then, for quantum optics,
by Glauber in 1963~\cite{Gla63}.
These states can be visualized in phase space by the Wigner function~\cite{Sch01,Gerry05book}. The concept of the coherent states has been extended to other dynamical group actions~\cite{Per86,Gaz09}.
Coherent states exhibit the Planck limit to phase space,
 known as the standard quantum limit or shot-noise limit for HW symmetry.

The coherent-state superpositions have been extensively studied for the harmonic oscillator~\cite{Gerry05book,PhysRevA.99.063813}, whose position and momentum operators obey $\mathfrak{hm}$(1) algebra for a single degree of freedom, and act on an infinite-dimensional Hilbert space. The phase-space features of a cat state
(superposition of two distinct coherent states~\cite{Mil86,YS86},
parameterized by mean particle or photon number~$\bar{n}$
corresponding to phase-space distance from the phase-space origin) are not limited in all phase-space
directions, and hence, cannot be considered as sub-Planck.

Compared to coherent states, 
cat states as a superposition of two coherent states with the same~$\bar{n}$ but opposite phases have $\sqrt{\bar{n}}$-enhanced sensitivity against displacements, with respect to specific directions in phase space.
Contrariwise, for a compass state,
this enhanced sensitivity to displacements is independent of phase-space directions. 
These same sub-Planck structures present in compass states appear in cat mixtures~\cite{Eff12}.
However, similar to cat states, the cat mixtures have shown this enhanced sensitivity along a specific direction in phase space.
These results have been generalized to the case of SU(2) dynamics~\cite{Naeem2021}.

Another symmetry of special interest for physicists is the Lie SU(1,1) group generated by the~$\mathfrak{su}$(1,1) algebra, which is associated with displacement-like operators and involves three generators as its basis elements, and acts on infinite-dimensional Hilbert space~\cite{Per86}.
The SU(1,1) coherent states focused on in this paper are those of Perelomov~\cite{Gerry1986,Ojeda2014}.

SU(1,1) symmetry is connected with many quantum optical systems~\cite{Wodkiewicz1985,Chaturvedi1991,Ban1992,Ban1993,ban1993b,Jing2011,Ojeda2014,Hudelist2014,Berrada2013,Yurke1896,Dong2016,Szigeti2017,barry2020,duan2022a,duan2022b}.
The bosonic or Schwinger realization of the~$\mathfrak{su}$(1,1) algebra has a connection to the squeezing properties of boson fields~\cite{Lo1993}.
For example, the single- and two-mode bosonic representations of the~$\mathfrak{su}$(1,1) algebra have immediate relevance to the single and two-mode squeezed states~\cite{Brif1996,Gerry1986,Gerry1991,Gerry1995,stoler1971,Yuen1976,Gerry01,yazdi2008,luc1992}, respectively. 

Similar to other dynamical groups, the SU(1,1) quasiprobability distributions defined over the hyperboloid surface are obtained through the Wigner function~\cite{Seyfarth2020,Klimov2021,Russel2021}.
The SU(1,1) Wigner distribution can  be visualized on the Poincar\'e\ disk by using the stereographic projection. The superposition of SU(1,1) coherent states has been extensively studied in different contexts~\cite{ban1994,Gerry1997,Miry2012}.
Moreover, the SU(1,1) coherent-state superpositions have shown strong nonclassical properties~\cite{Gerry1997}. 
Multiple proposals suggest ways to realize experimentally
SU(1,1) coherent-state superpositions ~\cite{Gerry1997,Miry2012,zheng2002generation}. Entangled coherent states~\cite{San92,San92E,San12}
have been extended to superpositions of SU(2) and SU(1,1) coherent states~\cite{WSP00}.

Here, we discuss the phase-space representations of interesting SU(1,1) coherent-state superpositions using the SU(1,1) Wigner function.
We show that by considering SU(1,1) coherent-state superpositions on the hyperboloid surface, one can build cat states, compass states, and cat-state mixtures. These quantum states have similar phase-space features as their counterparts of the HW and the SU(2) groups when represented on the Poincar\'e\ disk. For a coherent state, the Wigner distribution appears as a lobe at the location where it is pinned on the Poincar\'e\ disk.
The effective support of this lobe decreases as~$k$ ($k$ is the Bargmann index) grows, which scales as $\nicefrac1{k}$.
We show that the concept of sub-Planck structures is extended to the SU(1,1) group by associating the support area of the coherent states as the SU(1,1) counterpart
of the Planck action. In particular, we show that SU(1,1) compass state and cat-state mixtures have phase-space structures with extension proportional to $\nicefrac1{k}$ along any direction, which is a factor $\nicefrac1{\sqrt{k}}$ smaller than the extension found for coherent states. Two-mode bosonic realization of the $\mathfrak{su}$(1,1) algebra relates the Bargmann index $k$ to the asymmetry in photon numbers of correlated modes of two-mode squeezed number states. We then show that the existence of the sub-Planck structures are connected to larger values of this asymmetry in photon numbers of two correlated modes of the squeezed-number states. Moreover, we  verify that the SU(1,1) coherent-state superpositions of the present work have exactly the same enhanced sensitivity to
displacements found for their counterparts of the HW and the SU(2) groups, with the role of~$\bar{n}$ in the HW group and $j$ in the SU(2) group
played by~$k$ for the SU(1,1) groups.

Our paper is organized as follows.\;In~\S\ref{sec:background}, we review the basic concepts of sub-Planck structures associated with the~$\mathfrak{hw}(1)$ algebra. In~\S\ref{sec:su(1,1)_group}, we introduce specific SU(1,1) coherent-state superpositions and discuss their phase-space representation by the Wigner function. Here we explore various aspects associated with these states such as sub-Planck structures present in their Wigner function, two-photon realizations, and sensitivity against displacements. Furthermore, in this section we also compare the properties of these SU(1,1) states with their HW and SU(2) counterparts. Finally, we summarize our discussion in~\S\ref{sec:summary}.

\begin{figure*}[t]
\includegraphics[width=\textwidth]{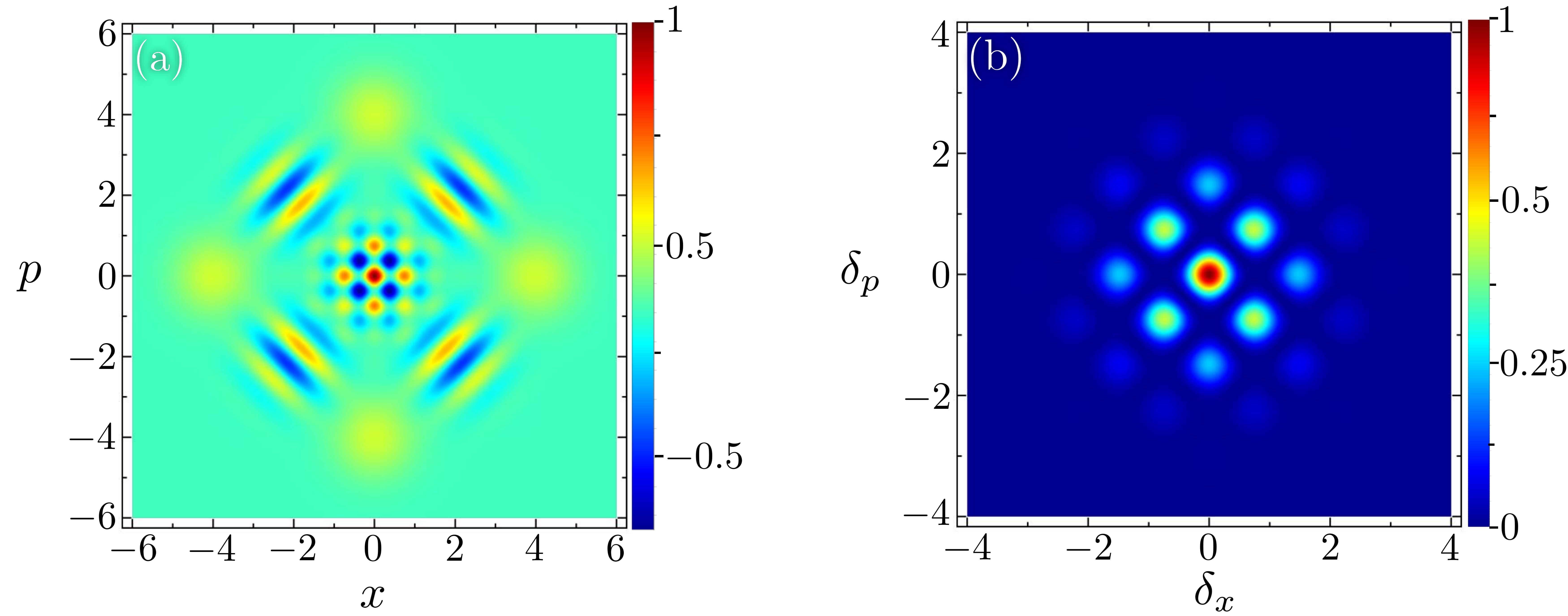}
\caption{For $x_0 = 4$,
(a)~Wigner function of the (Heisenberg-Weyl) compass state and (b)~overlap between compass state and its displaced versions. The quantities are in arbitrary units.}
\label{fig:figure1}
\end{figure*}
\section{Background}\label{sec:background}

In this section we present the background of the main concepts, including
phase-space representations of quantum states, sub-Planck structures, and the sensitivity of a quantum state against phase-space displacements. We explain these concepts by means of the compass state of the harmonic oscillator or~the $\mathfrak{hw}(1)$ algebra.
\subsection{Sub-Planck structures}\label{subsec:compass}

Let us start to explain some basics of the~$\mathfrak{hw}(1)$ algebra. This algebra is defined through the creation $\hat{a}^\dagger$ operator and anhilation operator $\hat{a}$ that satisfy the commutation relation $[\hat{a},\hat{a}^\dagger]=1$. The dimensionless versions 
of the position and momentum operators,
\begin{align}
    \hat{x}:=\frac{\hat{a}^\dagger+\hat{a}}{\sqrt{2}}\;\text{and}\;
    \hat{p}:=\frac{\text{i}(\hat{a}^\dagger-\hat{a})}{\sqrt{2}},
\end{align}
respectively,
obey the uncertainty relation $\Delta x\Delta p\geq1/2$, where
\begin{equation}
\label{eq:DeltaA2}
\Delta C^2:=\braket{\hat{C}^2}-\braket{\hat{C}}^2
\end{equation}
is the uncertainty of any operator~$\hat{C}$~\cite{Robertson1929, wheeler2014}.

A Schr\"{o}dinger coherent state is a nonspreading wave packet of the quantum harmonic oscillator~\cite{Sch35} and is defined as an eigenstate of the annihilation operator:
$\hat a\ket{\alpha}=\alpha\ket{\alpha}$ for $\alpha\in\mathbb{C}$~\cite{Gla63,Gerry05book}.
The coherent state is generated by displacing the vacuum state by
$\ket{\alpha}=\hat{D}(\alpha)\ket{0}$~\cite{Gaz09}
for $\hat{D}(\alpha):=\exp\left(\alpha\hat{a}^\dagger-\alpha^*\hat{a}\right)$.
In the Fock basis,
\begin{align}
	\ket{\alpha}=\text{e}^{-\frac{|\alpha|^2}{2}}\sum^{\infty}_{n=0}\frac{\alpha^n}{\sqrt{n!}}\ket{n},\,
	\{\ket{n};n\in\mathbb{N}\},
\end{align}
which yields a Poisson distribution of the particle number and mean~$\bar{n}=|\alpha|^2$.

The Wigner function for any arbitrary quantum state 
$\hat{\rho}$
is written as an expectation value of the parity kernel
as~\cite{CarlosNB15,Russel2021}
\begin{align}
    W_{\hat{\rho}}\left(\bm{r}\right)
    :=\text{tr}\left[\hat{\rho}\hat{\Pi}(\alpha)\right],\,
    \bm{r}:=(x,p)^\top,
\label{eq:wigner_general}
\end{align}
with
\begin{align}
 \hat{\Pi}(\alpha):=2\hat{D}(\alpha)\hat{\Pi}\hat{D}^\dagger(\alpha),\,
 \hat{\Pi}
:=\exp\left(\text{i}\pi\hat{a}^\dagger\hat{a}\right),
\end{align}
being the displaced parity operator.

Zurek~\cite{Zurek2001} showed that the compass state leads to sub-Planck structures in phase space,
and,
importantly,
these structures play a crucial role in enhancing its sensitivity towards phase-space displacements~\cite{Toscano06,Eff4,Naeem2021}.
The superposition of four coherent states with equal amplitude and maximally-spaced phases leading to the Zurek compass state (we omit state normalization throughout) is as follows:
\begin{align}\label{eq:compass_state}
    \ket{\psi}:=\ket{\nicefrac{x_0}{\sqrt{2}}}+\ket{\nicefrac{-x_0}{\sqrt{2}}}+\ket{\nicefrac{\text{i}x_0}{\sqrt{2}}}+\ket{\nicefrac{-\text{i}x_0}{\sqrt{2}}},
\end{align}
with $x_0 \in \mathbb{R}$. The Wigner function of this compass state is shown in Fig.~\ref{fig:figure1}(a). Throughout we normalize the Wigner functions with their maximum amplitudes $W_{\rho}(0)$.
The Wigner function of the compass state (\ref{eq:compass_state}) is written as a sum of the Wigner functions of the underlying coherent states plus the interferences between them, that is,
\begin{align}
W_{\ket{\psi_{\text{C}}}}(\bm{r})=W_{\text{coh}}(\bm{r})+I_\text{osc}(\bm{r})+I_\boxplus(\bm{r}),
\label{eq8}
\end{align}
where
\begin{equation}
W_{\text{coh}}(\bm{r})
:=\frac14\left[\text{e}^{-p^2}G(x;x_0)
+\text{e}^{-x^2}G(p;x_0)\right],
\end{equation}
with
\begin{equation}
\label{eq:doublepeak}
G(x;x_0)
:=\text{e}^{-\left(x-x_0\right)^2}
+\text{e}^{-\left(x+x_0\right)^2},
\end{equation}
represents the Wigner functions of four coherent states underlying the compass state. The locations of these coherent states in phase space are understood as the geographical points (east, west, north and south).
The second term in Eq.~(\ref{eq8}) is
\begin{equation}
    I_\text{osc}(\bm{r}):=\frac12\sum_{m_1,m_2=\pm 1}V(m_1 x,m_2 p),
\end{equation}
with
\begin{equation}
    V(x,p):=\text{e}^{-\left[\left(x-\frac{x_0}{2}\right)^2+\left(p-\frac{x_0}{2}\right)^2\right]}\cos\Big[x_0\left(x+p-\frac{x_0}{2}\right)\Big],
\end{equation}
and represents the quantum interference between northeast, northwest, southeast, and southwest pairs of the coherent states.

For our purpose, we focus on the central chessboardlike pattern
\begin{align}
 I_\boxplus(x,p):=\frac12\text{e}^{-(x^2+p^2)}\Big[\cos(2x_0 x)+\cos(2x_0 p)\Big].
\end{align}
This pattern
contains tiles of alternating sign
(denoted by different colors
in the Fig.~\ref{fig:figure1}(a)) with areas proportional to $x_0^{-2}$, and, hence, below than the Planck scale for $x_0\gg1$. The same sub-Planck structures  appear for the cat-state mixture~\cite{Eff12,Naeem2021}. 
Cat states do not have sub-Planck structures because the effective support of their phase-space structures appearing in the
interference pattern is limited only in the specific direction. The sensitivity of the compass state to displacements is discussed next.

\subsection{Sensitivity to displacements}\label{subsec:hw_sensitivity}
The sensitivity of any arbitrary quantum state $\hat{\rho}$ to displacements is obtained by calculating its overlap with their displaced versions $\hat{D}(\delta\alpha)\hat{\rho}\hat{D}^\dagger(\delta\alpha)$. This overlap is
\begin{equation}
\label{eq:overlap_HW}
\mathcal{F}_{\hat{\rho}}(\delta\alpha)
:=\text{tr}\big[\hat{\rho}\hat{D}(\delta\alpha)\hat{\rho}\hat{D}^\dagger(\delta\alpha)\big]	
=\left|\braket{\psi|\hat{D}(\delta\alpha)|\psi}\right|^2,
\end{equation}
where $\delta\alpha\in\mathbb{C}$ is an arbitrary displacement and the last equality holds when the states are pure, $\hat{\rho}=\ket{\psi}\bra{\psi}$. This quantity provides a measure~\cite{Audenaert14} for the distinguishability of the state and its displaced version. The smaller
the displacement $\delta \alpha$ needs to be in order to bring the overlap
to zero, the more sensitive the state is said to be against displacements. 

For a coherent state $\ket{\alpha}$ the overlap (\ref{eq:overlap_HW}) is
\begin{align}
 \mathcal{F}_{\ket{\alpha}}(\delta \alpha)=\text{e}^{-\frac12\left|\delta \alpha\right|^2},
\end{align}
where the smallest noticeable displacement that vanishes this overlap is above the Planck scale, $|\delta \alpha|>1$.
This inequality implies that the sensitivity of a given coherent state is at the standard quantum limit.

Consider the compass state, for which the overlap (\ref{eq:overlap_HW}) under the assumption $x_0\gg1$ and $|\delta \alpha|\ll1$ is
\begin{align}
\mathcal{F}_{\ket{\psi}}(\delta\alpha)=\frac14 \mathrm{e}^{-\frac12|\delta\alpha|^2}\big[\cos\left(x_0 \delta_x\right)+\cos\left(x_0 \delta_p\right)\big]^2,
\end{align}
with
\begin{equation}
\delta\alpha
=\delta_x+\text{i}\delta_p,\hspace{3mm} \delta_j\in\mathbb{R}.
\end{equation}
We plot this overlap in Fig.~\ref{fig:figure1}(b), and it vanishes when either of the conditions
\begin{equation}\label{eq:compassov}
	\delta_x\pm\delta_p=\frac{2m+1}{x_0}\pi,~m\in\mathbb{Z},
\end{equation}
are fulfilled.
As illustrated in Fig.~\ref{fig:figure1}(b), the overlap vanishes for the displacements $|\delta \alpha|\sim x^{-1}_0$ and the arbitrary phase. Hence, as compared to coherent states, a compass state with excitations $\bar{n}$ (here $\bar{n}=\nicefrac{x^2_0}{2}$) has shown $\sqrt{\bar{n}}$-enhanced sensitivity against displacements of any arbitrary directions in phase space.

For cat states and their mixtures this enhancement only occurs for displacements of specific directions. Having the same sub-Planck structures in phase-space, cat-state mixtures are inferior for metrology of compass states, for which the quantum superposition play a crucial role. Hence, sub-Planck structures are not the sole reason for the remarkable sensibility of compass states against displacements~\cite{Toscano06}. 

The coherent-state superpositions of the SU(2) group are  well discussed~\cite{San89,SG14,davis2020,Huang15,Huang18,Maleki}.
SU(2) quasiprobability distributions defined over the unit sphere are obtained through the Wigner function~\cite{davis2020,Varilly89,Heiss00,Klimov17,Glaser20}. The concept of sub-Planck structures has been extended to the SU(2) group~\cite{Naeem2021}. 
Specifically, by restricting SU(2) coherent-state superpositions along the equator, one can build cat states, compass states, and cat-state mixtures that have similar Wigner interference patterns as their HW counterparts when represented in the stereographic plane, with the role of~$x_0$ being played by $\sqrt{j}$~\cite{Naeem2021,davis2020}. The compass state and cat-state mixtures of the SU(2) group have shown the structures limited in all directions of the stereographic plane. These structures have an extension proportional to $\nicefrac1{j}$ in any direction, which is a factor $\nicefrac1{\sqrt{j}}$ smaller than the extension found for coherent states.
Furthermore, SU(2) cat states have shown an interference pattern
with structures limited only in one direction, just like their
HW counterparts. These states have exactly the same enhanced sensitivity to displacements found for the~$\mathfrak{hw}$(1) algebra, where $j$ has played the role of~$\bar{n}$~\cite{Naeem2021}.

\section{Sub-Planck structures of SU(1,1)}\label{sec:su(1,1)_group}
Different algebras are associated with different systems~\cite{Per86}.\;For example, the~$\mathfrak{hw}(1)$ algebra, discussed in the previous section, acts on an infinite-dimensional Hilbert space and is typically associated
with one-dimensional mechanical systems. Most recently SU(1,1) has achieved special attention for metrology~\cite{Jing2011,Hudelist2014,Berrada2013,Yurke1896,Dong2016,Szigeti2017,barry2020}. The SU(1,1) representation is associated with the Hamiltonian involving squeezing~\cite{Knight1987}. This is because the SU(1,1) displacement operator is considered as a squeezing operator, resulting in the SU(1,1) coherent state actually being the squeezed state~\cite{Brif1996,Gerry1986,Gerry1991,Gerry1995,stoler1971,Yuen1976,Gerry01,yazdi2008,luc1992}. In fact, squeezing of quantum states represents the leading strategy for enhanced quantum metrology~\cite{Maccone2004,drummond2004quantum}. It is interesting to extend the results found for one
algebra to different ones. It may provide a way to
devise experimental implementations of essentially any
algebra we are interested in. We have found it useful to generalize the concept of sub-Planck structures from the harmonic oscillator to the SU(2) group~\citep{Naeem2021}. In this spirit, in the following we generalize the concept of sub-Planck structures to the SU(1,1) group.

In~\S\ref{subsec:preliminary}, we review the main properties of the~$\mathfrak{su}$(1,1) algebra. In~\S\ref{subsec:su(1,1)_wigner}, we evaluate the Wigner function of the SU(1,1) coherent states. In~\S\ref{subsec:su(1,1)_superpositions}, we introduce specific coherent-state superpositions
and show how the concept of sub-Planck structures can be extended to the SU(1,1) case. In~\S\ref{subsec:two_photon}, we explore two-photon representations of these superpositions, and then in ~\S\ref{subsec:su(1,1)orthogonality}, we discuss their sensitivity against phase-space displacements. In~\S\ref{subsec:comparision}, we compare the properties of these SU(1,1) coherent-state superpositions with their HW and SU(2) counterparts.

 \begin{figure*}[t]
\centering
\includegraphics[width=0.97\textwidth]{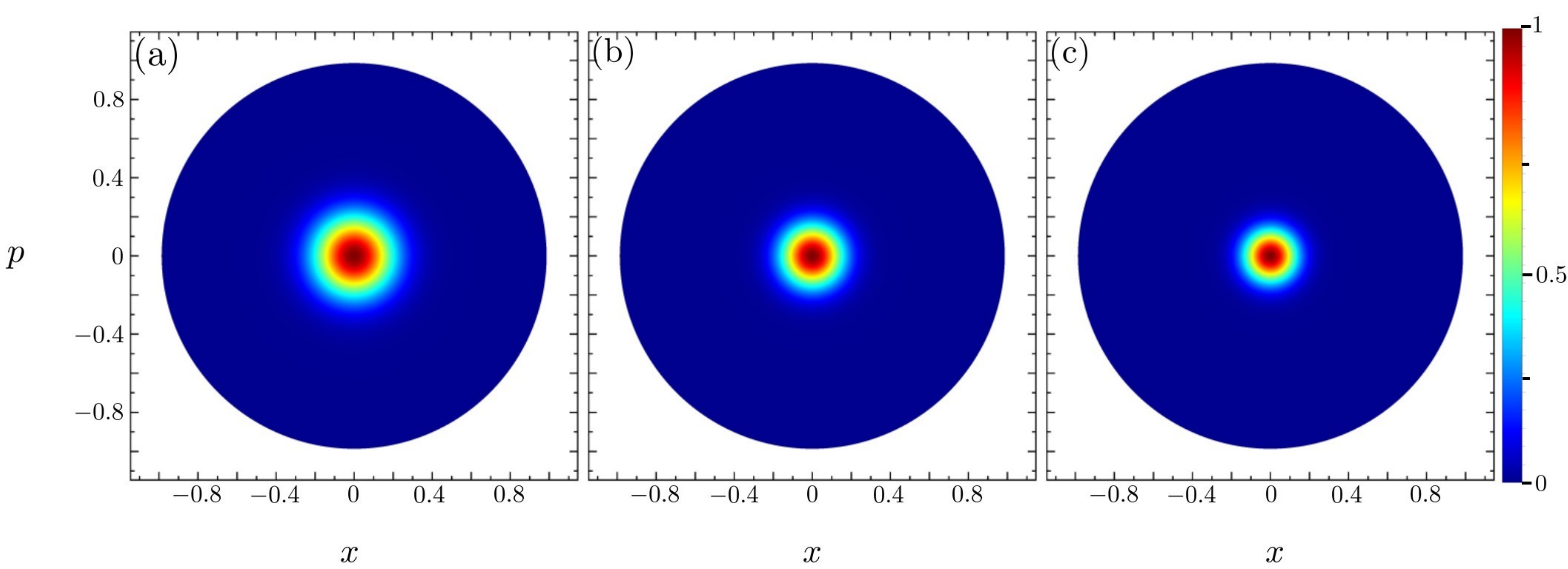}
\caption{Plots of the SU(1,1) Wigner function of the reference state $\ket{k,0}$ on the Poincar\'e\ disk. (a)~$k=6$, (b)~$k=10$, and (c) $k=14$. The quantities are in arbitrary units.}
\label{fig:figure2}
\end{figure*}
\subsection{Basics of SU(1,1)}\label{subsec:preliminary}

The~$\mathfrak{su}$(1,1) algebra is spanned by the generators $\hat{K}_+$, $\hat{K}_-$, and $\hat{K}_0$, which satisfy the commutation relations~\cite{Per86}
\begin{align}\label{eq:commutations_su(1,1)}
	\big[\hat{K}_0,\hat{K}_{\pm}\big]=\pm \hat{K}_{\pm},\,\big[\hat{K}_-,\hat{K}_+\big]=2\hat{K}_0.
\end{align}
These generators are written in terms of the Hermitian operators $\hat{K_1}$ and $\hat{K_2}$ as
\begin{align}
\hat{K}_{\pm}=\pm \text{i} (\hat{K}_1\pm \text{i}\hat{K}_2).
\end{align}
The action of the SU(1,1) generators on the Fock space states~$\{\ket{k,n};n\in\mathbb{N}\}$ satisfies
\begin{align}
\label{eq:commu_su(1,1)}
&\hat{K}_0\ket{k,n}=(k+n)\ket{k,n},\\
&\hat{K}_{+}\ket{k,n}=\sqrt{(n+1)(2k+n)}\ket{k,n+1},\\
&\hat{K}_{-}\ket{k,n}=\sqrt{n(2k+n+1)}\ket{k,n-1},
\end{align}
where $\ket{k,0}$ is the normalized reference state. For any irreducible representation the Casimir operator satisfies
\begin{align}
\hat{K}^2=&\hat{K}_0-\frac12(\hat{K}_{+}\hat{K}_{-}+\hat{K}_{-}\hat{K}_{+}),\\=&\hat{K}^2_0-\hat{K}^2_1-\hat{K}^2_2=k(k-1)\mathbb{1},
\label{eq:casimor_su(1,1)}
\end{align}
where~$k$ is the Bargmann index which separates different irreducible
representations. We restrict to positive discrete series for which $k>0$.

The SU(1,1) displacement operator admits either of the forms~\cite{mufti1993,fujii2001introduction,ban1993b}
\begin{equation}
	\hat{D}(\zeta):=\text{e}^{\xi \hat{K}_+-\xi^* \hat{K}_-}
	=\text{e}^{\zeta \hat{K}_+}\text{e}^{\ln(1-\left|\zeta\right|^2) \hat{K}_0}\text{e}^{-\zeta^* \hat{K}_-},
\label{eq:displacement_su11_dis}
\end{equation}
where
\begin{equation}
\xi=\frac{\tau}{2}\text{e}^{\text{i}\varphi},\,\zeta=\text{e}^{\text{i}\varphi}\tanh\left(\frac{\tau}{2}\right),
\end{equation}
with $-\infty<\tau<\infty$ and $0<\varphi<2\pi$. 
The parameter $|\zeta|$ is restricted to the interior of the Poincar\'e\
disk, $0\le|\zeta|<1$, whereas $\xi$ is defined on the upper sheet of the two-sheet hyperboloid surface. 

Multiplication of two operators obey~\cite{Per86}
\begin{equation}
	\hat{D}(\zeta_ 1)\hat{D}(\zeta_ 2)
	=\hat{D}(\zeta_ 3)\text{e}^{\text{i}\phi \hat{K}_0},
	\label{eq:product_operator}
\end{equation}
where
\begin{equation}
	\zeta_3=\frac{\zeta_1+\zeta_2}{1+\zeta^*_1\zeta_2},\phi=-2\arg(1+\zeta^*_1\zeta_2).
\end{equation}
As Eq.~(\ref{eq:product_operator}) is independent of~$k$, 
this is proved by setting
\begin{equation}
\hat{K}_0=\frac{\sigma_3}2,\, \hat{K}_{1,2}=\frac{\text{i}\sigma_{1,2}}2,
\end{equation}
for the Pauli matrices $\{\sigma_j\}$.
\subsection{SU(1,1) coherent states and the Wigner function}\label{subsec:su(1,1)_wigner}
SU(1,1) coherent states are obtained by displacing~(\ref{eq:displacement_su11_dis}) the reference state $\ket{k,0}$
according to~\cite{Gerry1991,Gerry1986}
\begin{align}
	\ket{\zeta}=\hat{D}(\zeta)\ket{k,0}=(1-\left|\zeta\right|^2)^k\sum^{\infty}_{n=0}\sqrt{\frac{\Gamma(n+2k)}{n!\Gamma(2k)}}\zeta^n\ket{k,n}.
\label{eq:coherentstate_su11}
\end{align}
Equivalently, SU(1,1) coherent states can  be associated with points on the two-sheeted hyperboloid surface through the hyperbolic version of the Bloch vector,
namely,
\begin{equation}
    \bm{n}=(\cosh\tau,\sinh\tau\cos\varphi,\sinh\tau\sin\varphi).
\end{equation}
Using Eqs.~(\ref{eq:product_operator})
and~(\ref{eq:coherentstate_su11}),
the overlap between any two arbitrary coherent states is
\begin{equation}\label{eq:overlap_SU(11)}
	\braket{\zeta_1|\zeta_2}=\left[\frac{(1-|\zeta_1|^2)(1-|\zeta_2|^2)}{\left(1-\zeta^*_1 \zeta_2\right)^2}\right]^k.
\end{equation}
The overlap between a reference state $\ket{k,0}$ and any state $\ket{\zeta}$ is approximately Gaussian as a function of~$\zeta$:
\begin{align}
\braket{k,0|\zeta}=(1-|\zeta|^2)^k\approx \mathrm{e}^{-k|\zeta|^2},\,k\gg1.
\end{align}
The width of this overlap is proportional to $\nicefrac1{\sqrt{k}}$, which decreases as~$k$ grows.

Similarly, the overlap between two coherent states located at different points of the hyperboloid surface decreases with~$k$ according to
\begin{align}
	|\braket{\zeta_1|\zeta_2}|=\cosh^{-k}(\chi/2),
\end{align}
where
\begin{equation}
	\cosh\chi=\cosh\tau_1\cosh\tau_2-\cos(\varphi_1-\varphi_2)\sinh \tau_1 \sinh \tau_2,
\end{equation}
with
\begin{equation}
\zeta_j=\exp\left(\text{i}\varphi_j\right)\tanh(\tau_j/2).
\end{equation}

The SU(1,1) Wigner function of operator~$\hat{\rho}$ is evaluated via the expectation value of the displaced parity operator~\cite{Seyfarth2020,Klimov2021,Russel2021},
\begin{align}
	W_{\hat{\rho}}\left(\zeta\right):=\text{tr}\big[\hat{\rho}\hat{\Pi}(\zeta)\big],
	\label{eq:wigner_su(1,1)}
\end{align}
where
\begin{equation}
	\hat{\Pi}(\zeta):=\hat{D}(\zeta) \hat{\Pi}\hat{D}^{\dagger}(\zeta),
\end{equation}
and $\hat{\Pi}:=\exp\left[\text{i}\pi(\hat{K}_0-k)\right]$ is the parity operator for SU(1,1)~\cite{edwin2018}.
The SU(1,1) Wigner distribution is visualized on the surface of the Poincar\'e\ disk via stereographic projection
\begin{align}
\zeta:=x+\text{i}p.
\end{align}
The unit disk can be lifted to the upper sheet of the two-sheeted hyperboloid by the inverse
stereographic map.


Using Eq.~(\ref{eq:wigner_su(1,1)}), the Wigner function of the operator $\ket{\zeta_1}\bra{\zeta_2}$ is easily found to be
\begin{widetext}
\begin{align}\label{eq:su(1,1)_coh_general}
	W_{\ket{\zeta_2}\bra{\zeta_1}}(\zeta)=\Bigg[\frac{\left(|\zeta|^2-1\right)^2\left(|\zeta_1|^2-1\right)\left(|\zeta_2|^2-1\right)\left(\zeta_2\zeta^*-1\right)\left(\zeta_1^*\zeta-1\right)}{(\zeta_1\zeta^*-1)(\zeta_2^*\zeta-1)\left(1-2\zeta\zeta_1^*+\zeta_2\zeta_1^*+\zeta_2\zeta_1^*+|\zeta|^2-2\zeta^*\zeta_2+|\zeta|^2\zeta_1^*\zeta_2\right)^2}\Bigg]^k \text{e}^{2 \text{i}k \text{arg}\left(\frac{1-\zeta_1 \zeta^*}{1-\zeta_2 \zeta^*}\right)}.
\end{align}
\end{widetext}
The detailed derivation of this Wigner function is provided in the Appendix.

In the particular case of the reference state $\ket{0}=\ket{k,0}$, the Wigner function is obtained
\begin{equation}
W_{\ket{0}}(\zeta)=\left(\frac{|\zeta|^2-1}{|\zeta|^2+1}\right)^{2k}.
\end{equation}
The corresponding Wigner distribution, which we show in Fig.~\ref{fig:figure2},
appears as a lobe. The support area of this lobe decreases isotropically as $k$ grows. We can approximate this Wigner function as a Gaussian of the form $\exp\left(-4k |\zeta|^2\right)$
for $k\gg1$.
Hence, the extension of the SU(1,1) coherent state along any direction in phase space is proportional to $\nicefrac1{\sqrt{k}}$, which is precisely the same scaling that we found for the width of the overlap between coherent states.
In the following, we show that the notion of sub-Planck structures is extended to the SU(1,1) group by associating the effective support of the SU(1,1) coherent state as a counter part of the SU(1,1) Planck action.

\begin{figure*}
\includegraphics[width=0.99\textwidth]{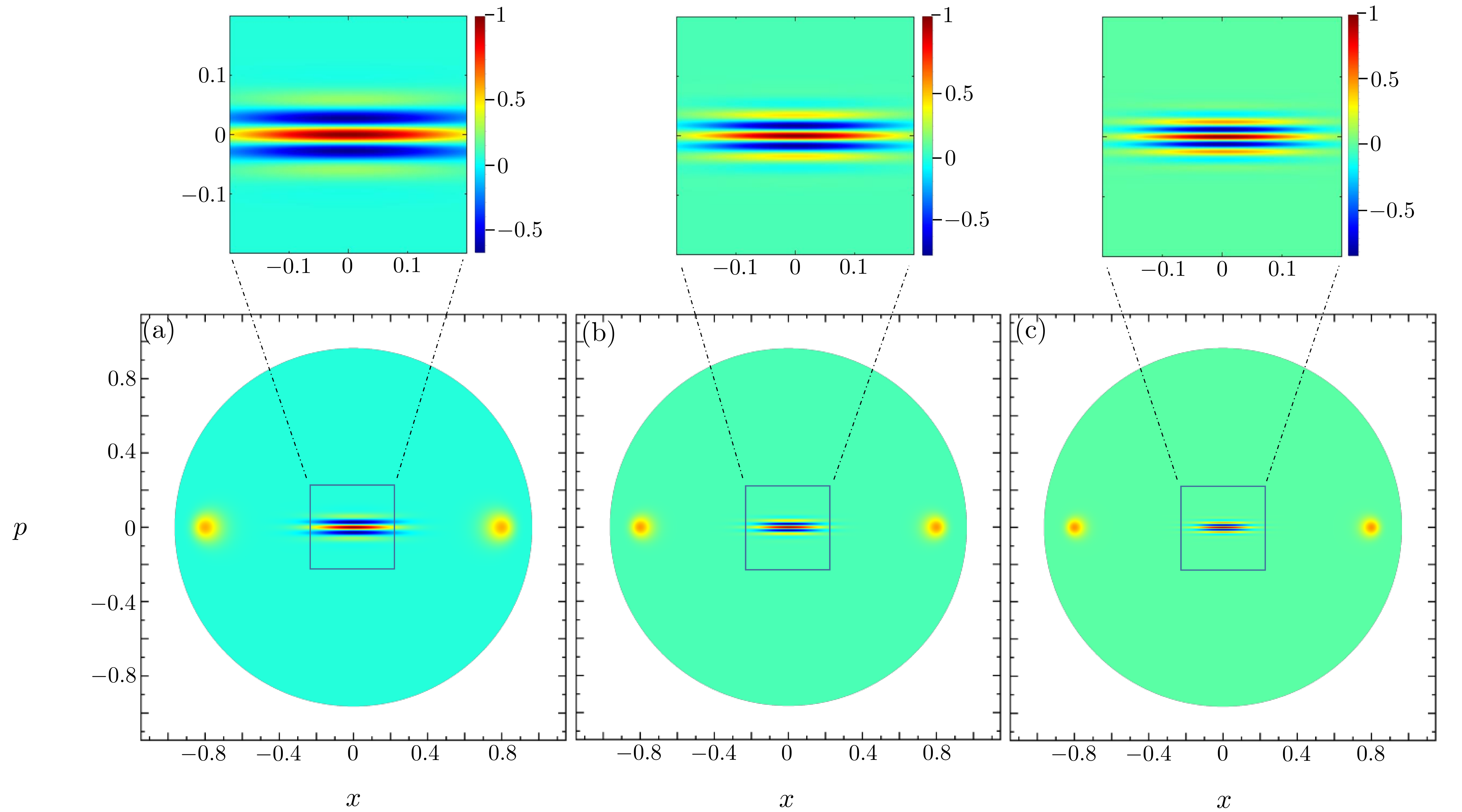}
\caption{Plots of the SU(1,1) Wigner function of the cat state on the Poincar\'e\ disk: (a)~$k=6$, (b)~$k=10$, and (c) $k=14$. Insets represent the interference pattern of each case. In all cases $\zeta_0=0.8$. The quantities are in arbitrary units.}
\label{fig:figure3}
\end{figure*}
\subsection{SU(1,1) coherent-state superpositions}
\label{subsec:su(1,1)_superpositions}
The SU(1,1) cat states (superposition of two distinct coherent states) have been discussed ~\cite{ban1994,Gerry1997,Miry2012,Klimov2021}.
In particular,
the ‘horizontal’ cat typically refers
to the superposition of coherent states along the horizontal axis of the Poincar\'e\ disk:
\begin{equation}\label{eq:horizontalcat_SU(1,1)}
	\ket{\psi_\text{H}}:=\ket{\zeta_0}+\ket{-\zeta_0},
\end{equation}
where $\zeta_0\in \mathbb{R}$. The corresponding Wigner function of this cat state is
\begin{align}
W_{\ket{\psi_{\text{H}}}}(\zeta)=W_{\ket{\zeta_0}}(\zeta)+W_{\ket{-\zeta_0}}(\zeta)+I_{\text{H}}(\zeta),
\end{align}
where each term is fairly easy to obtain using Eq.~(\ref{eq:su(1,1)_coh_general}). The first two terms represent the Wigner functions of the coherent states,
\begin{equation}
W_{\ket{\pm\zeta_0}}(\zeta)=\frac12\left[\frac{(\zeta_0^2-1)(|\zeta|^2-1)}{(\zeta_0^2+1)(|\zeta|^2+1)\pm 2\zeta_0 (\zeta+\zeta^*)}\right]^{2k},
\end{equation}
\begin{figure*}[t]
\includegraphics[width=0.99\textwidth]{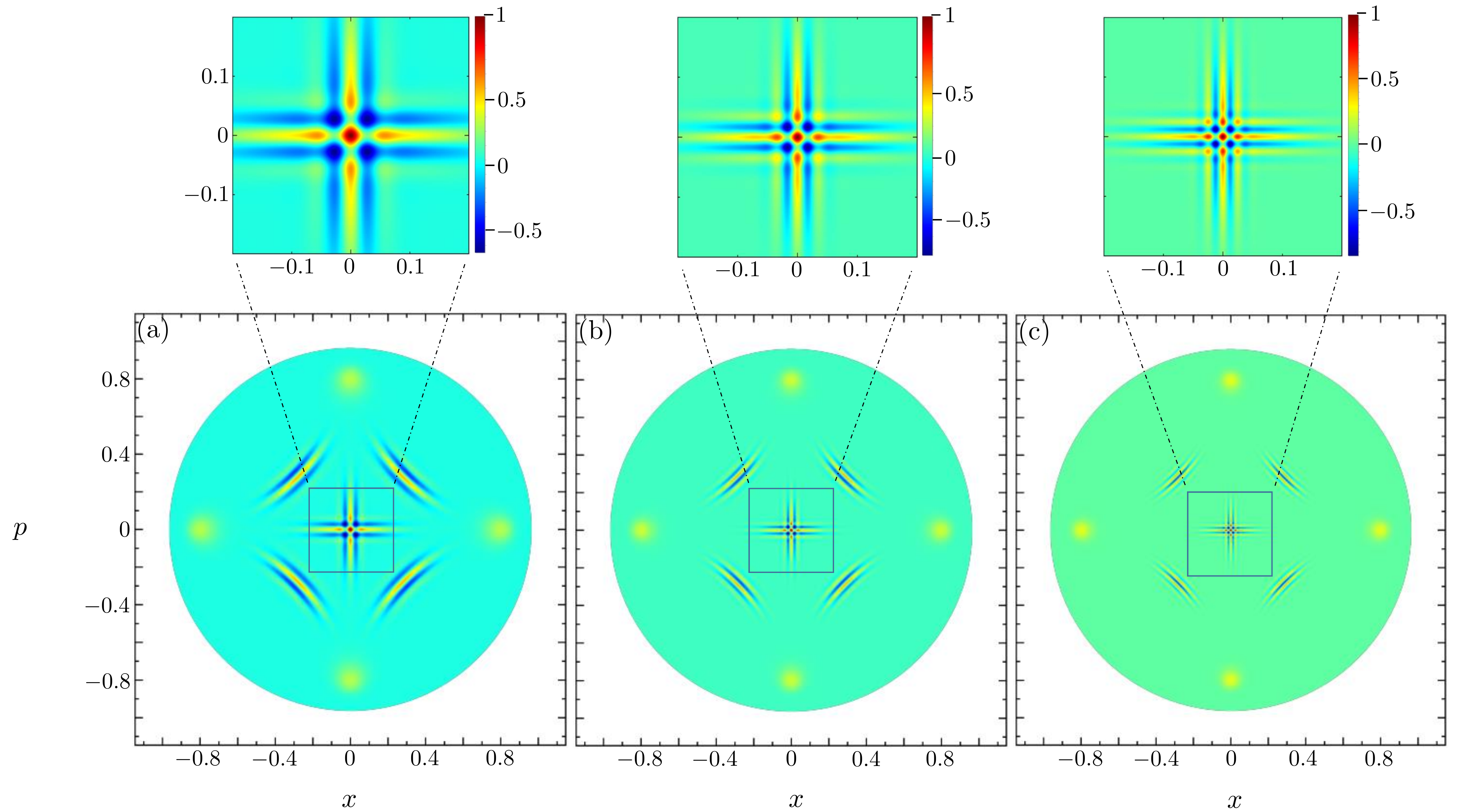}
\caption{Plots of the SU(1,1) Wigner function of the compass state on the Poincar\'e\ disk: (a)~$k=6$, (b)~$k=10$, and (c) $k=14$. Insets represent the central interference pattern of each case. In all cases $\zeta_0=0.8$. The quantities are in arbitrary units.}
\label{fig:figure4}
\end{figure*}
and the last term provides the interference between the underlying coherent states
\begin{widetext}
\begin{align}\label{eq:catinter}
I_{\text{H}}(\zeta):=\left[\frac{(\zeta^2_0-1)^2(|\zeta|^2-1)^2}{1-2(2\zeta^2+1)\zeta^2_0+\zeta_0^4+(\zeta^2_0-1)^2|\zeta|^4+2\zeta (\zeta^2_0+1)^2\zeta^*-4\zeta^2_0\zeta^*}\right]^k \cos\left[2k\arg(\Theta)\right],
\end{align}
\end{widetext}
with 
\begin{align}
 \Theta=\frac{(1-\zeta_0\zeta^*)(\zeta_0\zeta-1)}{(1+\zeta_0\zeta^*)(\zeta_0\zeta-1)+(\zeta+\zeta_0)(\zeta_0-\zeta^*)}.   
\end{align}

In Fig.~\ref{fig:figure3} we plot the corresponding Wigner function on the Poincar\'e\ disk. Two lobes appear on the unit disk at the locations $(\pm\zeta_0,0)$ are representing the coherent states.
In addition, interference appears as an oscillatory pattern directed along the $p$ direction of the stereographic plane. 
As illustrated in Fig.~\ref{fig:figure3} this interference pattern becomes pronounced (i.e., the number of oscillation increases) as the representation index~$k$ increases.

Along the $p$ axis ($x = 0$) the interference (\ref{eq:catinter}) becomes
\begin{align}
 I_{\text{H}}(p)=\left[\frac{(\zeta^2_0-1)^2(p^2-1)^2}{(\zeta^2_0-1)^2(1+p^4)+2p^2(\zeta^4_0+6\zeta^2_0+1)}\right]^k\cos\left(\theta^\prime \right),
\end{align}
where
\begin{align}\label{eq:thetalabel}
\theta^\prime=2k\tan^{-1}\left(\frac{4\zeta_0 p}{\zeta^2_0-1}\right).
\end{align} 
The zeros of the interference pattern $I_{\text{H}}(\zeta)$ occur when
\begin{align}
p=\pm \frac{(\zeta^2_0-1)}{4\zeta_0}\tan\left[\frac{(2m+1)\pi}{4k}\right],\,m\in\mathbb{Z}.
\end{align}
This means that the first zeros are
located at
\begin{equation}
p=\pm \frac{\zeta^2_0-1}{4\zeta_0}\tan\left[\frac{\pi}{4k}\right]\approx \pm \frac{\zeta^2_0-1}{16\zeta_0 k}\pi,\,k\gg1.
\end{equation}
Hence, the extension of the interference patches along the $p$ direction is proportional to $\nicefrac1{k}$ for $k\gg1$. In contrast, along the $x$ axis ($p=0$),
interference is simply approximated by
\begin{align}
I_{\text{H}}(x)=&\left(\frac{x^2-1}{x^2+1}\right)^{2k}\approx \text{exp}(-4kx^2),\,k\gg1.
\end{align}
Therefore, along the $x$ direction the extension of the interference pattern is proportional to $\nicefrac1{\sqrt{k}}$. 
This is precisely the same extension that we found for a coherent state along the $x$ direction.
Support of interference structures of the SU(1,1) horizontal cat state is limited only along the vertical direction of the phase space.

Similarly, we can  build cat states along the vertical axis of the
stereographic plane as
\begin{align}
	\ket{\psi_{\text{V}}}:=\ket{\text{i}\zeta_0}+\ket{-\text{i}\zeta_0},
\end{align}
whose Wigner function appears to be the same as one of the horizontal cat
states, but rotated by $\nicefrac{\pi}{2}$ in the Poincar\'e\ disk, that is,
\begin{align}
	W_{\ket{\psi_\text{V}}}(\zeta)=&\nonumber W_{\ket{\text{i}\zeta_0}}(\zeta)+W_{\ket{-\text{i}\zeta_0}}(\zeta)+I_\text{V}(\zeta),
	\\
	=& W_{\ket{\psi_\text{H}}}(p+\text{i}x).
	\end{align}
Here,
$W_{\ket{\pm\text{i}\zeta_0}}(\zeta)=W_{\ket{\pm \zeta_0}}(p+\text{i}x)$ represents the Wigner
function of underlying coherent states,
and $I_\text{V}(\zeta)=I_\text{H}(p+\text{i}x)$ is the interference.

\begin{figure*}[t]
\includegraphics[width=0.99\textwidth]{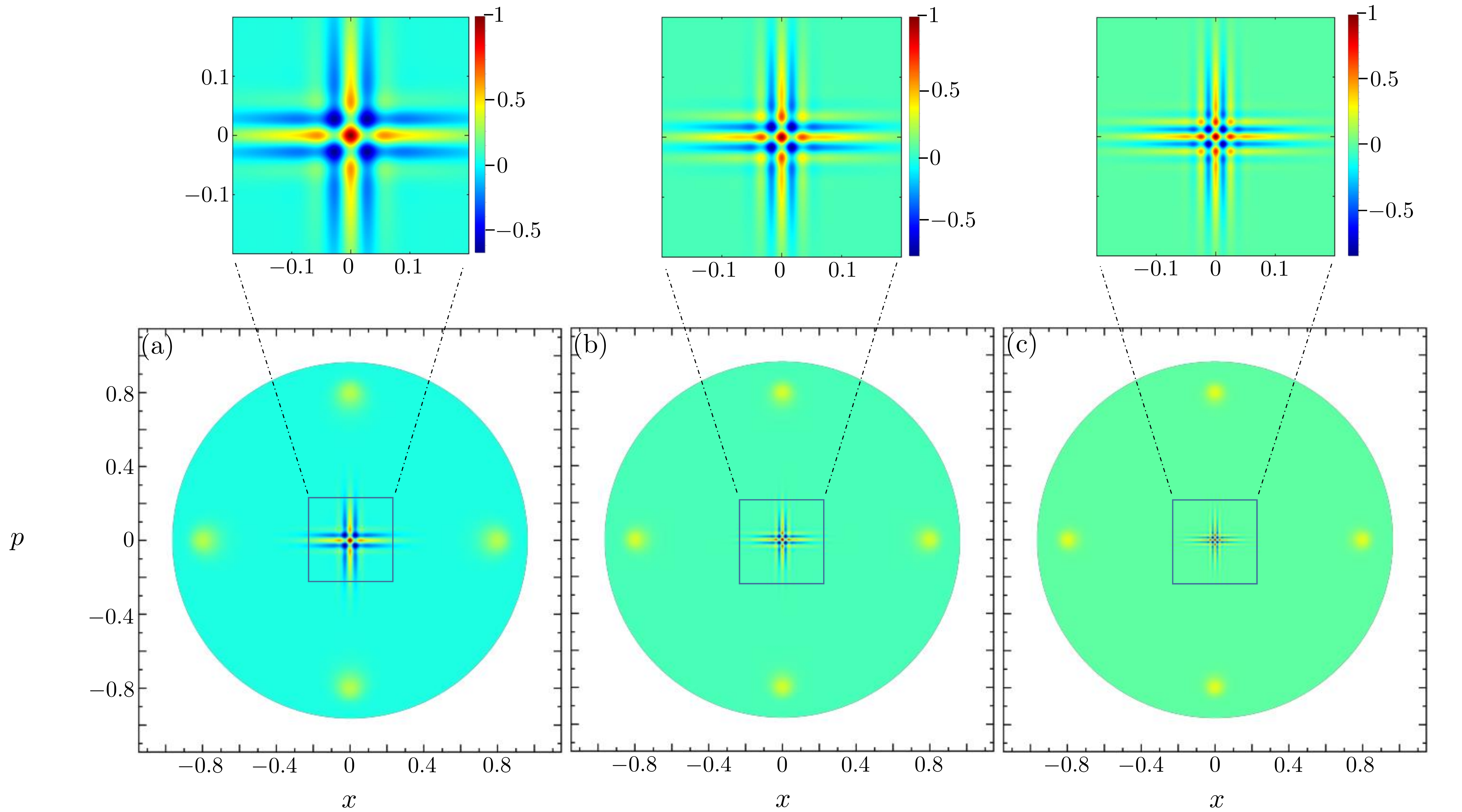}
\caption{Plots of the SU(1,1) Wigner function of cat-state mixtures on the Poincar\'e\ disk: (a)~$k=6$, (b)~$k=10$, and (c) $k=14$. Insets represent the central interference pattern of each case. In all cases $\zeta_0=0.8$. The quantities are in arbitrary units.}
\label{fig:figure5}
\end{figure*}

Let us consider now the superposition of horizontal and
vertical cat states, leading to the SU(1,1) compass state
\begin{align}\label{eq:compass_su(1,1)}
\ket{\psi_{\text{C}}}:=\ket{\psi_{\text{H}}}+\ket{\psi_{\text{V}}}.
\end{align}
The corresponding Wigner function is shown in
Fig.~\ref{fig:figure4}. This Wigner function is written as a sum of the Wigner functions of individual cat states plus the interference between these (cat-like interference patterns located at the
northeast, northwest, southeast, and southwest positions). We can clearly see four lobes centered at positions
$\left(\pm\zeta_0,0\right)$ and $\left(0,\pm\zeta_0\right)$ on the the Poincar\'e\ disk, which correspond to the coherent states.
Note that,
for $k\gg1$,
the chessboardlike pattern around the origin of the Poincar\'e\ disk is evident. The support area of a fundamental tile appears in a chessboardlike pattern that decreases isotropically in phase space as $k$ increases.

We focus on this central interference pattern, which is written as the sum of the interferences of the horizontal and vertical cat states,
that is,
\begin{align}
	I_\boxplus(\zeta)=I_\text{H}(\zeta)+I_\text{V}(\zeta).
\end{align}
The extension of each tile in this pattern is proportional to $\nicefrac1{k}$ along any arbitrary direction in phase space, which is a factor $\nicefrac1{\sqrt{k}}$ smaller than the extension found for coherent states. These results show that, as promised, the concept of sub-Planck structures is generalized to the SU(1,1) group.

These sub-Planck structures present by the mixture of two cat states. In particular, we consider the incoherent mixture of horizontal and vertical cat states
\begin{align}
	\label{eq:catmixture_SU(1,1)}
	\hat{\rho}_\text{M}:=\ket{\psi_{\text{H}}}\bra{\psi_{\text{H}}}+\ket{\psi_{\text{V}}}\bra{\psi_{\text{V}}},
\end{align}
whose Wigner function is equal to the sum of the Wigner functions of horizontal and vertical cat states, that is, 
\begin{equation}
W_{\hat{\rho}_\text{M}}(\zeta)=W_{\ket{\psi_{\text{H}}}}(\zeta)+W_{\ket{\psi_{\text{V}}}}(\zeta).
\end{equation}
This Wigner function is shown in Fig.~\ref{fig:figure5}, where the chessboardlike pattern appears around the origin of the Poincar\'e\ disk. Hence,
the same chessboardlike pattern with sub-Planck structures  appears for the SU(1,1) cat-state mixtures.

\begin{figure*}[t]
\centering
\includegraphics[width=\textwidth]{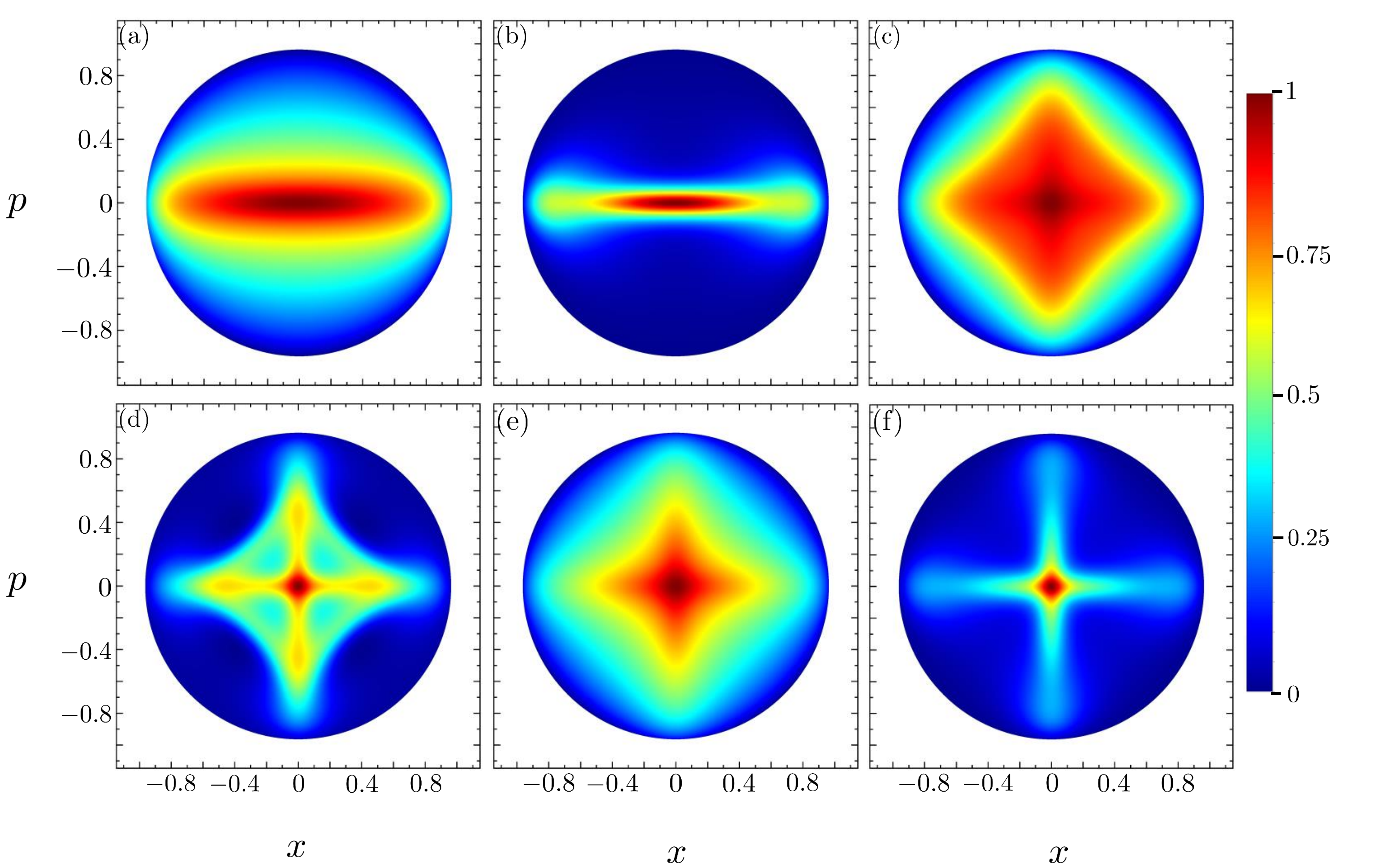}
\centering
\caption{The Wigner functions of the  SU(1,1) coherent-state superpositions considered in this work are shown on the Poincar\'e\ disk.  Panels (a)~and (b)~represent the Wigner functions of the cat state with $k=\nicefrac14$ and $k=\nicefrac{3}{4}$, respectively. Similarly, panels (c) and (d) shows the Wigner functions of the compass state for $k=\nicefrac14$ and $k=\nicefrac{3}{4}$, respectively. The Wigner functions of cat-state mixtures for $k=\nicefrac14$ and $k=\nicefrac{3}{4}$ are shown by panels (e) and (f), respectively. In all cases $\zeta_0=0.8$. The quantities are in arbitrary units.}
\label{fig:figure6}
\end{figure*}

\subsection{Correlated coherent states of the SU(1,1) group}
\label{subsec:two_photon}
In this subsection, we review the relation between a few well known quantum states and coherent
states associated with SU(1,1) group. The relevance of the~$\mathfrak{su}(1,1)$ algebra to the physical system can be obtained through the realization of the generators in terms of the operators of the underlying physical system. Here we focus on the bosonic realizations of the~$\mathfrak{su}(1,1)$ algebra corresponding to one and two modes~\cite{Brif1996,Gerry1986,Gerry1991,Gerry1995,stoler1971,Yuen1976,Gerry01,yazdi2008,luc1992}.

As a first example, we consider a single boson-mode system in which the elements of the~$\mathfrak{su}(1,1)$ algebra are expressed as a single set of boson annihilation and creation operators as
\begin{align}\label{eq:onemode_su11}
    \hat{K}_+=\frac12\hat{a}^{\dagger2},\, \hat{K}_-=\frac12\hat{a}^2,\,\hat{K}_0=\frac14\left(\hat{a}\hat{a}^{\dagger}+\hat{a}^{\dagger}\hat{a}\right).
\end{align}
The Casimir operator is $\hat{K}^2=-\nicefrac{3}{16}$, which leads to the Bargmann indices $k=\nicefrac14$ and $k=\nicefrac34$.
For these irreducible representations, the SU(1,1) displacement operator~(\ref{eq:displacement_su11_dis}) is
identified as a one-mode squeezed operator. The representation associated with the Bargmann index $k=\nicefrac{1}{4}$ 
is the even-numbered Fock states, while for $k=\nicefrac{3}{4}$ only the odd-numbered Fock states~\cite{Gerry01}.
Hence, for $k=\nicefrac14$,
the SU(1,1) Perelomov coherent state can be considered as an ordinary squeezed-vacuum state
\begin{align}
 \ket{\zeta,\nicefrac14}=(1-\left|\zeta\right|^2)^{\nicefrac14}\sum^{\infty}_{n=0}\frac{\sqrt{2n!}}{2^n n!}\zeta^n\ket{2n}.
\label{eq:coherentstate_su11squeezed}
\end{align}
For $k=\nicefrac{3}{4}$, the corresponding SU(1,1) Perelomov coherent state is just the squeezed one-photon state~\cite{Gerry01}
\begin{align}
 \ket{\zeta,\nicefrac{3}{4}}=(1-\left|\zeta\right|^2)^{\nicefrac34}\sum^{\infty}_{n=0}\frac{\sqrt{(2n+1)!}}{2^n n!}\zeta^n\ket{2n+1}.
\label{eq:coherentstate_su11squeezed2}
\end{align}
Hence, for these irreducible representations, the SU(1,1) coherent-state superpositions of the present work can be taken as the superpositions of the ordinary squeezed vacuum and squeezed one-photon states. In Fig.~\ref{fig:figure6} we plot the corresponding Wigner functions of each case. For the given parameters these Wigner functions appear as positive peaked distributions.
The superpositions of ordinary squeezed-vacuum states have been investigated~\cite{TANG201586,barry1989,quantumrep2021,Happ2018,quesene}.

Now we briefly review the realization two-mode standard case. Let ($\hat{a}_1$, $\hat{a}_2$) and ($\hat{a}_1^\dagger$, $\hat{a}_2^\dagger$) be, respectively, the annihilation and creation operators of modes 1 and 2. Furthermore, let $\ket{n_1}$ and $\ket{n_2}$ represent the number states of these two modes, and the complete number-state basis of the two-mode field is
\begin{align}
\ket{n_1,n_2}=\ket{n_1}\otimes \ket{n_2}.
\end{align}
The $\mathfrak{su}(1,1)$ algebra can  be realized by two-mode annihilation and creation operators as
\begin{align}
 & \hat{K}_+=\hat{a}_1^\dagger \hat{a}_2^\dagger,~\hat{K}_-=\hat{a}_1 \hat{a}_2,\nonumber\\
 &\hat{K}_0=\frac12\left(\hat{a}_1^\dagger \hat{a}_1+\hat{a}_2^\dagger \hat{a}_2+1\right).
\end{align}
These SU(1,1) operators obey the commutation relations~(\ref{eq:commutations_su(1,1)}), and their action on the two-mode states can be described as
\begin{align}
&\hat{K}_0\ket{n_1,n_2}=\frac12\left(n_1+n_2+1\right)\ket{n_1,n_2},\\&\hat{K}_+\ket{n_1,n_2}=\sqrt{(n_1+1)(n_2+1)}\ket{n_1+1,n_2+1},\\&\hat{K}_-\ket{n_1,n_2}=\sqrt{n_1 n_2}\ket{n_1-1,n_2-1}.
\label{eq:commu_twomode}
\end{align}
The Casimir operator (\ref{eq:casimor_su(1,1)}) in this case becomes
\begin{align}
\hat{K}^2_0=\frac14\left(\Delta^2-1\right),
\end{align}
where
\begin{align}
\Delta= \hat{a}_1^\dagger \hat{a}_1-\hat{a}_2^\dagger \hat{a}_2,
\end{align}
whose eigenvalue is equal to the
difference between the number of quanta in modes 1 and 2, i.e., $n_1-n_2$.

The representations that we obtain are those for which this difference is constant.
The $\mathfrak{su}$(1,1) basis~$\ket{k,n}$ can be identified by
\begin{align}\label{eq:irr}
k=\frac12(q+1),\, n=\frac12\left(n_1+n_2-q\right),~q=0,1,2,\dots
\end{align}
where~$q$ is the degeneracy parameter representing the eigenvalue of~$|\Delta|$, and it measures asymmetry in the photon number of two correlated modes.
We assume that mode~1 has~$q$ more photons than mode~2, so that $n_1=n_2+q$ and $n=n_2=0,1,2,\dots$.
Thus, the weight states of SU(1,1) becomes $\ket{k,n}=\ket{n_2,n_2+q}$, with $n=0,1,2,\dots$ or,
more conveniently,
it can be just simply written as  $\ket{n_2,n_2+q}=\ket{n,n+q}$ (with $n_2=n$). Therefore, SU(1,1) Perelomov coherent states (\ref{eq:coherentstate_su11}) can be written in terms of the two-mode squeezed number states as
\begin{align}
    \ket{\zeta,q}=(1-|\zeta|^2)^{\nicefrac{1+q}{2}}\sum^\infty_{n=0}\sqrt{\frac{(n+q)!}{n! q!}}\zeta^n \ket{n,n+q}.
    \label{eq:coherent_state_twophotons}
\end{align}
Note that the state $\ket{q,0}$ will be interpreted as the ground state of the relevant unitary irreducible representations of SU(1,1).
For $q=0$ we just have the familiar two-mode squeezed vacuum state. For $q>0$ it is the state obtained by the action of the two-mode squeezed vacuum operator on the number state $\ket{q,0}$. Hence, SU(1,1) coherent-state superpositions considered in this work can be just superpositions of ordinary two-mode squeezed number states. 
Superpositions of ordinary two-mode squeezed vacuum states have been investigated in Ref.~\cite{Cardoso2021}. 

The photon-number distribution of the SU(1,1) coherent states (\ref{eq:coherent_state_twophotons}) appears as a Poissonian distribution for $q=0$ (zero fluctuations in system)~\cite{Gerry1991,Gerry91b}. However, as~$q$ grows, the distribution has a peak value at $n>0$. Higher values of~$q$ inject more photons in the system. Hence, this distribution is sub-Poissonian as~$q$ increases.
As mentioned earlier, larger values of the Bargmann index~$k$ yields the sub-Planck structures in the phase space of compass states. Note that $k$ relates to the degeneracy parameter~$q$ by Eq.~(\ref{eq:irr}). In other words, we can say that the sub-Planck structures of the compass states are associated with $q\gg1$ for two-boson-mode standard case of the SU(1,1). This can be understood in a way similar to that for compass states of the harmonic oscillator, i.e., injecting more photons in the states brings more sub-Planckness in the phase space. The influence of squeezing on the quantum decoherence that occurs in a two-qubit system has been investigated~\cite{Basit2021}. The $\mathfrak{su}$(1,1) algebra can also be associated with the four-mode boson field by a four-boson realization of SU(1,1)~\cite{PhysRevA.63.042310}.

\begin{figure*}[t]
\centering
\includegraphics[width=0.99\textwidth]{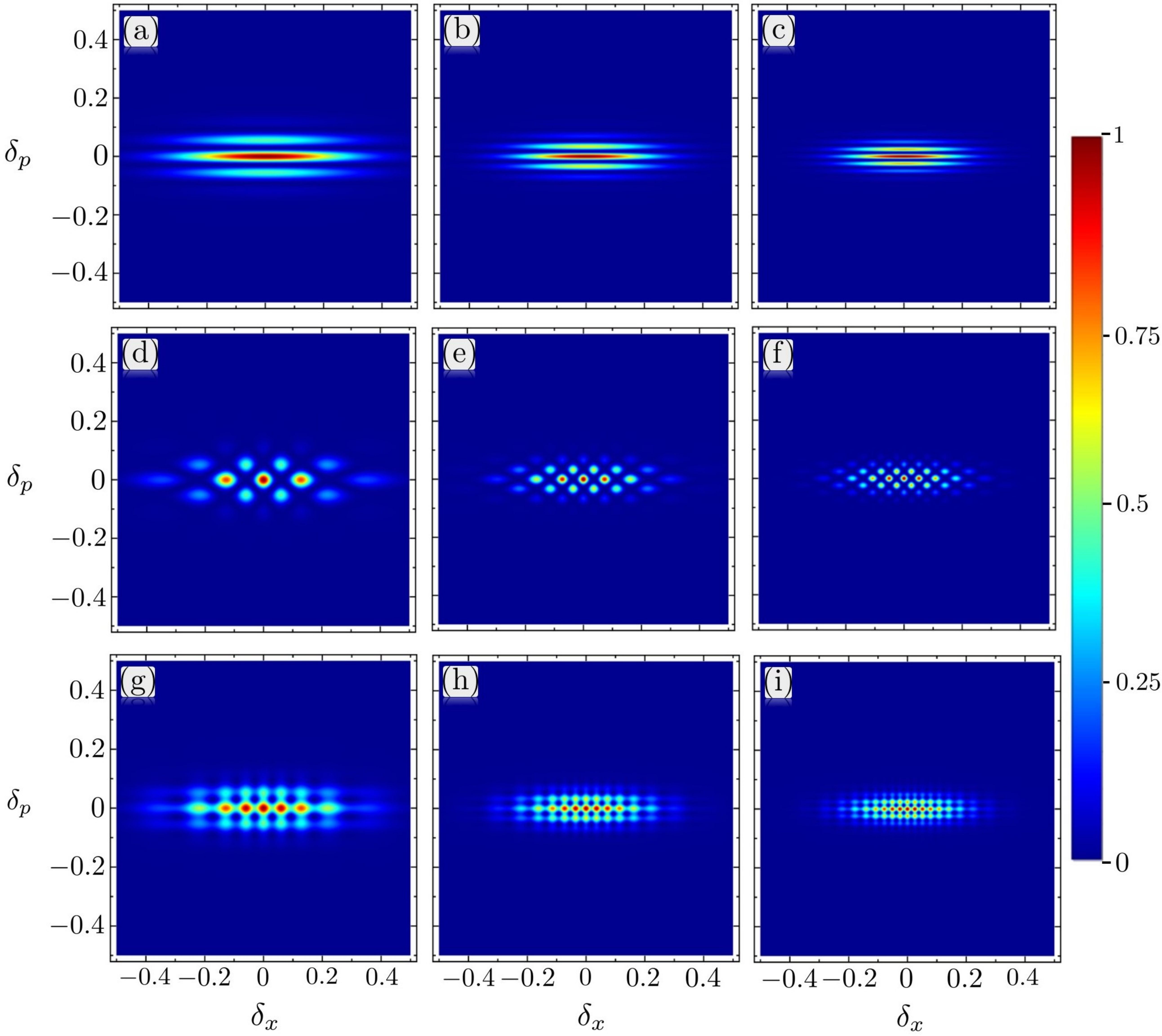}
\centering
\caption{Overlap between SU(1,1) states considered in this work and their slightly displaced versions. The left column corresponds to $k=6$, the middle column to $k=10$, and the right column to $k=14$. Panels (a)-(c) represent cat-state overlaps, panels (d)-(f) represent the overlaps of the compass state, and panels (g)-(i) are the overlaps of the cat-state mixtures. In all cases $\zeta_0=0.8$. The quantities are in arbitrary units.}
\label{fig:figure7}
\end{figure*}

\subsection{Sensitivity against SU(1,1) displacements}\label{subsec:su(1,1)orthogonality}

In this subsection, we discuss the sensitivity against phase-space displacements of SU(1,1) coherent-state superpositions presented in the preceding section. We compute the overlap between the states and their $\delta\zeta$-displaced versions, as given by Eq.~(\ref{eq:overlap_HW}).

Let us first consider SU(1,1) coherent states. We already discussed in~\S\ref{subsec:preliminary}, for $k\gg1$, this overlap is approximated in Gaussian form as $\exp\left(-k |\delta\zeta|^2\right)$.
Hence, the sensitivity to displacements for SU(1,1) coherent states scales as $\nicefrac1{\sqrt{k}}$.
In the following we have to compare the sensitivity of SU(1,1) coherent-state superpositions with this scaling.

Consider next the horizontal cat state (\ref{eq:horizontalcat_SU(1,1)}). The overlap (\ref{eq:overlap_HW}) for this state under the approximation $k\gg1$ leads to
\begin{align}\label{eq:overlap_horizontalcat_su(1,1)}
	\mathcal{F}_{\ket{\psi_\text{H}}}\left(\delta\zeta\right)=\frac12\left[\frac{(\zeta_0^2-1)^2(|\delta\zeta|^2-1)}{1-2\zeta_0^2+4\zeta_0^2\delta_p^2+\zeta_0^4}\right]^{2k}\cos^2(2k\theta),
\end{align}
where
\begin{align}
\theta=\tan^{-1}\left(\frac{2\zeta_0 \delta_p}{\zeta^2_0-1}\right)
\end{align}
and
\begin{equation}
	\delta\zeta:=\delta _x+\text{i}\delta_ p,
\end{equation}
with $\delta_j\in\mathbb{R}$. Note, for $k\gg1$ and $|\delta\zeta|\ll1$, the contribution of the cross terms between the
coherent states to the overlap, e.g., $\bra{\zeta_0}\hat{D}(\delta\zeta)\ket{-\zeta_0}$ and $\bra{-\zeta_0}\hat{D}(\delta\zeta)\ket{\zeta_0}$, is negligible.
The condition to make this overlap equal to zero is
\begin{align}
\delta _p=&\frac{\zeta^2_0-1}{2\zeta_0}
\tan\left[\frac{(2m+1)\pi}{4k}\right]\\\approx&\frac{(\zeta^2_0-1)(2m+1)}{8\zeta_0k}\pi,\,m\in\mathbb{Z},
k\gg1.
\end{align}
Thus, for large~$k$, the displacement $\delta_ p \sim \nicefrac1{k}$ along the vertical direction of the stereographic plane can make the horizontal cat state orthogonal. This overlap is plotted for different values of~$k$ in Figs.~\ref{fig:figure7}(a)-\ref{fig:figure7}(c). As compared to the coherent state, the horizontal cat state shows $\nicefrac1{\sqrt{k}}$ times higher sensitivity against displacements. This occurs
for displacements in the vertical direction in the stereographic plane. However, for the horizontal displacement it does not show the enhanced sensitivity compared to coherent states.

For $k\gg1$, the overlap (\ref{eq:overlap_HW}) of the compass state (\ref{eq:compass_su(1,1)}) is
\begin{align}\label{eq:overlap_compass_su(1,1)}
\mathcal{F}_{\ket{\psi_\text{C}}}(\delta\zeta)=\frac12\left[\sqrt{\mathcal{F}_{\ket{\psi_\text{H}}}\left(\delta\zeta\right)}+\sqrt{\mathcal{F}_{\ket{\psi_\text{V}}}\left(\delta\zeta\right)}\right]^2,
\end{align}
with
\begin{equation}
    \mathcal{F}_{\ket{\psi_\text{V}}}\left(\delta\zeta\right)=\mathcal{F}_{\ket{\psi_\text{H}}}\left(\delta\zeta_p+\text{i}\delta\zeta_x\right).
\end{equation}
We plot this overlap in Figs.~\ref{fig:figure7}(d)-\ref{fig:figure7}(f) with different values of~$k$.
This result shows that the SU(1,1) compass state has $\nicefrac1{\sqrt{k}}$ times higher sensitivity against displacements compared to the coherent states, but now  this enhanced sensitivity is independent of the displacement directions.

Finally, we consider the cat-state mixture~(\ref{eq:catmixture_SU(1,1)}). The overlap~(\ref{eq:overlap_HW}) for this state leads to be
\begin{align}
\mathcal{F}_{\hat{\rho}_{\text{M}}}\left(\delta\zeta\right)=\mathcal{F}_{\ket{\psi_\text{H}}}\left(\delta\zeta\right)+\mathcal{F}_{\ket{\psi_\text{V}}}\left(\delta\zeta\right).
\end{align}
This overlap is plotted for different values of~$k$ in Figs.~\ref{fig:figure7}(g)-\ref{fig:figure7}(i).
Now, the $\sqrt{k}$-enhanced sensitivity is present for displacements directed along the $\delta _x=\pm \delta _p$ directions.

\subsection{Analogies: SU(1,1) versus HW and SU(2)}
\label{subsec:comparision}
In this subsection, we compare properties (phase-space features and sensitivity against the displacements) of the SU(1,1) coherent-state superpositions of the present work with their HW and SU(2) counterparts. SU(1,1) compass states and cat-state mixtures have structures of extensions proportional to $\nicefrac1{k}$ ($k$ is the Bargmann index) in all phase-space directions, which is~$\nicefrac1{\sqrt{k}}$ times smaller than the extension found for coherent states. 
Interference features of SU(1,1) cat states are limited only in one direction, just like their HW and SU(2) counterparts. This shows that the Wigner functions of these states have exactly the same phase-space features as their HW and SU(2) counterparts when plotted on the Poincar\'e\ disk, with the role of~$x_0$ (distance of the coherent states from the origin) in the HW case and $\sqrt{j}$ ($j$ the angular momentum) in the SU(2) case
now played by $\sqrt{k}$.

These SU(1,1) coherent-state superpositions are shown be sensitive against displacements that are lower than the sensitivity found for the coherent states by a factor of $\nicefrac1{\sqrt{k}}$. This enhanced sensitivity for the SU(1,1) compass state is independent of the displacements directions in phase space. Whereas, for cat states and their mixtures,
this enhancement always occurs in specific directions. This shows that these SU(1,1) states have shown exactly
the same behavior against the displacements as their HW and SU(2) counterparts, with the role of~$\bar{n}$ in the HW case and $j$ in the SU(2) being played by~$k$ for SU(1,1).

\section{Summary}\label{sec:summary}
We have shown that by considering coherent-state superpositions on the hyperboloid surface, one can build SU(1,1) cat states, compass states, and cat-state mixtures with phase-space features similar to those of their HW and SU(2) counterparts when their Wigner functions are represented on the Poincar\'e\ disk. In particular, both SU(1,1) cat-state superpositions (compass state) and mixtures have sub-Planck structures in phase space, but interference structures of cat states are not considered to be sub-Planck (since they are not limited in all phase-space directions). Moreover, these SU(1,1) coherent state superpositions also behave similarly to their HW and SU(2)
counterparts regarding their sensitivity against displacements. This generalizes sub-Planck structures found in the HW and SU(2) cases to the SU(1,1) group. 

We have reviewed the two-mode bosonic realization of the $\mathfrak{su}$(1,1) algebra, which relates the Bargmann index $k$ to the asymmetry in photons numbers of correlated modes of two-mode squeezed number states. Then we have shown that the existence of the sub-Planck structures is associated with larger asymmetry in photon numbers of two correlated modes of the squeezed number state. In a similar way, the enhanced sensitivity of these superpositions can also be connected with this asymmetry in photon numbers of two modes i.e., higher asymmetry in photon numbers of these two modes corresponds to better enhanced sensitivity against the displacements.

A number of interesting schemes have been presented for the implementation of SU(1,1) cat states~\cite{Gerry1997,Miry2012,zheng2002generation}. Another future direction will concern how to generate the SU(1,1) compass states introduced in our work. Some of these schemes can be adapted to achieve the generation of SU(1,1) compass states, which otherwise will require a complete new proposal to generate superpositions of four coherent states.

\begin{acknowledgements}
The authors acknowledge support by the National Natural Science Foundation of China (Grants Nos.\ 11835011 and No.\ 12174346).
We thank Liwei Duan for insightful discussions.
\end{acknowledgements}
\appendix

\section{SU(1,1) Wigner function}\label{appendix:appendixA}
In this section, we provide the more detailed derivations of the Wigner function for SU(1,1) coherent states. For the operator $\hat{\rho}=\ket{\zeta_2}\bra{\zeta_1}$, we rewrite its Wigner function as
\begin{align}\label{eq:appendix_su(1,1)wig}
W_{\ket{\zeta_2}\bra{\zeta_1}}(\zeta)=&\nonumber\bra{\zeta_1}\hat{D}(\zeta) \hat{\Pi}\hat{D}^{\dagger}(\zeta)\ket{\zeta_2},\\=&\bra{k,0}\hat{D}^{\dagger}(\zeta_1)\hat{D}(\zeta) \hat{\Pi}\hat{D}^{\dagger}(\zeta)\hat{D}(\zeta_2)\ket{k,0}.
\end{align}
In some cases simpler expressions are found using the alternative form of the composition property of displacement operators, which we rewrite
here as
\begin{align}
	\hat{D}(\zeta_ 1)\hat{D}(\zeta_ 2)=\text{e}^{\text{-i}\phi \hat{K}_0}\hat{D}(\zeta_ 3),
	\label{eq:product2_operator}
\end{align}
with
\begin{equation}
\zeta_3=\frac{\zeta_1+\zeta_2}{1+\zeta_1\zeta^*_2},\, \phi=-2\arg(1+\zeta_1\zeta^*_2).
\end{equation}
Using composition laws given by Eqs.~(\ref{eq:product_operator}) and~(\ref{eq:product2_operator}), we simplify Eq.~(\ref{eq:appendix_su(1,1)wig}) as 
\begin{equation}
W_{\ket{\zeta_2}\bra{\zeta_1}}(\zeta)=\text{e}^{2\text{i}k\text{arg}\left(\frac{1-\zeta_1\zeta^*}{1-\zeta_2\zeta^*}\right)}\bra{\zeta^{\prime}_1}\hat{\Pi}\ket{\zeta_2{^\prime}},
\end{equation}
with~$\zeta^{\prime}_1=\frac{\zeta_1-\zeta}{1-\zeta_1\zeta^*},\, \zeta^{\prime}_2=\frac{\zeta_2-\zeta}{1-\zeta_2\zeta^*}.$
 This expression is easily simplified to obtain Eq.~(\ref{eq:su(1,1)_coh_general}).

\bibliography{References}

\begin{thebibliography}{109}%
\makeatletter
\providecommand \@ifxundefined [1]{%
 \@ifx{#1\undefined}
}%
\providecommand \@ifnum [1]{%
 \ifnum #1\expandafter \@firstoftwo
 \else \expandafter \@secondoftwo
 \fi
}%
\providecommand \@ifx [1]{%
 \ifx #1\expandafter \@firstoftwo
 \else \expandafter \@secondoftwo
 \fi
}%
\providecommand \natexlab [1]{#1}%
\providecommand \enquote  [1]{``#1''}%
\providecommand \bibnamefont  [1]{#1}%
\providecommand \bibfnamefont [1]{#1}%
\providecommand \citenamefont [1]{#1}%
\providecommand \href@noop [0]{\@secondoftwo}%
\providecommand \href [0]{\begingroup \@sanitize@url \@href}%
\providecommand \@href[1]{\@@startlink{#1}\@@href}%
\providecommand \@@href[1]{\endgroup#1\@@endlink}%
\providecommand \@sanitize@url [0]{\catcode `\\12\catcode `\$12\catcode
  `\&12\catcode `\#12\catcode `\^12\catcode `\_12\catcode `\%12\relax}%
\providecommand \@@startlink[1]{}%
\providecommand \@@endlink[0]{}%
\providecommand \url  [0]{\begingroup\@sanitize@url \@url }%
\providecommand \@url [1]{\endgroup\@href {#1}{\urlprefix }}%
\providecommand \urlprefix  [0]{URL }%
\providecommand \Eprint [0]{\href }%
\providecommand \doibase [0]{https://doi.org/}%
\providecommand \selectlanguage [0]{\@gobble}%
\providecommand \bibinfo  [0]{\@secondoftwo}%
\providecommand \bibfield  [0]{\@secondoftwo}%
\providecommand \translation [1]{[#1]}%
\providecommand \BibitemOpen [0]{}%
\providecommand \bibitemStop [0]{}%
\providecommand \bibitemNoStop [0]{.\EOS\space}%
\providecommand \EOS [0]{\spacefactor3000\relax}%
\providecommand \BibitemShut  [1]{\csname bibitem#1\endcsname}%
\let\auto@bib@innerbib\@empty
\bibitem [{\citenamefont {Robertson}(1929)}]{Robertson1929}%
  \BibitemOpen
  \bibfield  {author} {\bibinfo {author} {\bibfnamefont {H.~P.}\ \bibnamefont
  {Robertson}},\ }\bibfield  {title} {\bibinfo {title} {The uncertainty
  principle},\ }\href {https://doi.org/10.1103/PhysRev.34.163} {\bibfield
  {journal} {\bibinfo  {journal} {Phys. Rev.}\ }\textbf {\bibinfo {volume}
  {34}},\ \bibinfo {pages} {163} (\bibinfo {year} {1929})}\BibitemShut
  {NoStop}%
\bibitem [{\citenamefont {Wheeler}\ and\ \citenamefont
  {Zurek}(1983)}]{wheeler2014}%
  \BibitemOpen
  \bibfield  {author} {\bibinfo {author} {\bibfnamefont {J.~A.}\ \bibnamefont
  {Wheeler}}\ and\ \bibinfo {author} {\bibfnamefont {W.~H.}\ \bibnamefont
  {Zurek}},\ }\href@noop {} {\emph {\bibinfo {title} {Quantum Theory and
  Measurement}}}\ (\bibinfo  {publisher} {Princeton University Press},\
  \bibinfo {address} {Princeton, NJ},\ \bibinfo {year} {1983})\BibitemShut
  {NoStop}%
\bibitem [{\citenamefont {Wigner}(1932)}]{Wig32}%
  \BibitemOpen
  \bibfield  {author} {\bibinfo {author} {\bibfnamefont {E.}~\bibnamefont
  {Wigner}},\ }\bibfield  {title} {\bibinfo {title} {On the quantum correction
  for thermodynamic equilibrium},\ }\href
  {https://doi.org/10.1103/PhysRev.40.749} {\bibfield  {journal} {\bibinfo
  {journal} {Phys. Rev.}\ }\textbf {\bibinfo {volume} {40}},\ \bibinfo {pages}
  {749} (\bibinfo {year} {1932})}\BibitemShut {NoStop}%
\bibitem [{\citenamefont {Weyl}(1950{\natexlab{a}})}]{Wey50}%
  \BibitemOpen
  \bibfield  {author} {\bibinfo {author} {\bibfnamefont {H.}~\bibnamefont
  {Weyl}},\ }\href@noop {} {\emph {\bibinfo {title} {The Theory of Groups and
  Quantum Mechanics}}}\ (\bibinfo  {publisher} {Dover},\ \bibinfo {address}
  {New York},\ \bibinfo {year} {1950})\BibitemShut {NoStop}%
\bibitem [{\citenamefont {Moyal}(1949)}]{Moy49}%
  \BibitemOpen
  \bibfield  {author} {\bibinfo {author} {\bibfnamefont {J.~E.}\ \bibnamefont
  {Moyal}},\ }\bibfield  {title} {\bibinfo {title} {Quantum mechanics as a
  statistical theory},\ }\href {https://doi.org/10.1017/S0305004100000487}
  {\bibfield  {journal} {\bibinfo  {journal} {Math. {P}roc. {C}ambridge
  {P}hilos. Soc.}\ }\textbf {\bibinfo {volume} {45}},\ \bibinfo {pages} {99}
  (\bibinfo {year} {1949})}\BibitemShut {NoStop}%
\bibitem [{\citenamefont {Gerry}\ and\ \citenamefont
  {Knight}(2005)}]{Gerry05book}%
  \BibitemOpen
  \bibfield  {author} {\bibinfo {author} {\bibfnamefont {C.}~\bibnamefont
  {Gerry}}\ and\ \bibinfo {author} {\bibfnamefont {P.}~\bibnamefont {Knight}},\
  }\href@noop {} {\emph {\bibinfo {title} {Introductory Quantum Optics}}}\
  (\bibinfo  {publisher} {Cambridge University Press},\ \bibinfo {address}
  {England, Cambridge},\ \bibinfo {year} {2005})\BibitemShut {NoStop}%
\bibitem [{\citenamefont {Schleich}(2001)}]{Sch01}%
  \BibitemOpen
  \bibfield  {author} {\bibinfo {author} {\bibfnamefont {W.~P.}\ \bibnamefont
  {Schleich}},\ }\href@noop {} {\emph {\bibinfo {title} {Quantum Optics in
  Phase Space}}}\ (\bibinfo  {publisher} {Wiley-VCH},\ \bibinfo {address}
  {Weinheim},\ \bibinfo {year} {2001})\BibitemShut {NoStop}%
\bibitem [{\citenamefont {Drummond}\ and\ \citenamefont
  {Ficek}(2004)}]{drummond2004quantum}%
  \BibitemOpen
  \bibfield  {author} {\bibinfo {author} {\bibfnamefont {P.~D.}\ \bibnamefont
  {Drummond}}\ and\ \bibinfo {author} {\bibfnamefont {Z.}~\bibnamefont
  {Ficek}},\ }\href@noop {} {\emph {\bibinfo {title} {Quantum {S}queezing}}}\
  (\bibinfo  {publisher} {Springer-Verlag},\ \bibinfo {address} {Berlin},\
  \bibinfo {year} {2004})\BibitemShut {NoStop}%
\bibitem [{\citenamefont {Zurek}(2001)}]{Zurek2001}%
  \BibitemOpen
  \bibfield  {author} {\bibinfo {author} {\bibfnamefont {W.~H.}\ \bibnamefont
  {Zurek}},\ }\bibfield  {title} {\bibinfo {title} {Sub-\uppercase{P}lanck
  structure in phase space and its relevance for quantum decoherence},\ }\href
  {https://doi.org/10.1038/35089017} {\bibfield  {journal} {\bibinfo  {journal}
  {Nature}\ }\textbf {\bibinfo {volume} {412}},\ \bibinfo {pages} {712}
  (\bibinfo {year} {2001})}\BibitemShut {NoStop}%
\bibitem [{\citenamefont {Zurek}(2003)}]{Zurek2003}%
  \BibitemOpen
  \bibfield  {author} {\bibinfo {author} {\bibfnamefont {W.~H.}\ \bibnamefont
  {Zurek}},\ }\bibfield  {title} {\bibinfo {title} {Decoherence, einselection,
  and the quantum origins of the classical},\ }\href
  {https://doi.org/10.1103/RevModPhys.75.715} {\bibfield  {journal} {\bibinfo
  {journal} {Rev. Mod. Phys.}\ }\textbf {\bibinfo {volume} {75}},\ \bibinfo
  {pages} {715} (\bibinfo {year} {2003})}\BibitemShut {NoStop}%
\bibitem [{\citenamefont {Toscano}\ \emph {et~al.}(2006)\citenamefont
  {Toscano}, \citenamefont {Dalvit}, \citenamefont {Davidovich},\ and\
  \citenamefont {Zurek}}]{Toscano06}%
  \BibitemOpen
  \bibfield  {author} {\bibinfo {author} {\bibfnamefont {F.}~\bibnamefont
  {Toscano}}, \bibinfo {author} {\bibfnamefont {D.~A.~R.}\ \bibnamefont
  {Dalvit}}, \bibinfo {author} {\bibfnamefont {L.}~\bibnamefont {Davidovich}},\
  and\ \bibinfo {author} {\bibfnamefont {W.~H.}\ \bibnamefont {Zurek}},\
  }\bibfield  {title} {\bibinfo {title} {Sub-\uppercase{P}lanck phase-space
  structures and \uppercase{h}eisenberg-limited measurements},\ }\href
  {https://doi.org/10.1103/PhysRevA.73.023803} {\bibfield  {journal} {\bibinfo
  {journal} {Phys. Rev. A}\ }\textbf {\bibinfo {volume} {73}},\ \bibinfo
  {pages} {023803} (\bibinfo {year} {2006})}\BibitemShut {NoStop}%
\bibitem [{\citenamefont {Dalvit}\ \emph {et~al.}(2006)\citenamefont {Dalvit},
  \citenamefont {de~Matos~Filho},\ and\ \citenamefont {Toscano}}]{Eff4}%
  \BibitemOpen
  \bibfield  {author} {\bibinfo {author} {\bibfnamefont {D.~A.~R.}\
  \bibnamefont {Dalvit}}, \bibinfo {author} {\bibfnamefont {R.~L.}\
  \bibnamefont {de~Matos~Filho}},\ and\ \bibinfo {author} {\bibfnamefont
  {F.}~\bibnamefont {Toscano}},\ }\bibfield  {title} {\bibinfo {title} {Quantum
  metrology at the \uppercase{h}eisenberg limit with ion trap motional compass
  states},\ }\href {https://doi.org/10.1088/1367-2630/8/11/276} {\bibfield
  {journal} {\bibinfo  {journal} {New J. Phys.}\ }\textbf {\bibinfo {volume}
  {8}},\ \bibinfo {pages} {276} (\bibinfo {year} {2006})}\BibitemShut {NoStop}%
\bibitem [{\citenamefont {Jacquod}\ \emph {et~al.}(2002)\citenamefont
  {Jacquod}, \citenamefont {Adagideli},\ and\ \citenamefont
  {Beenakker}}]{Eff1}%
  \BibitemOpen
  \bibfield  {author} {\bibinfo {author} {\bibfnamefont {P.}~\bibnamefont
  {Jacquod}}, \bibinfo {author} {\bibfnamefont {I.}~\bibnamefont {Adagideli}},\
  and\ \bibinfo {author} {\bibfnamefont {C.~W.~J.}\ \bibnamefont {Beenakker}},\
  }\bibfield  {title} {\bibinfo {title} {Decay of the {L}oschmidt {E}cho for
  {Q}uantum {S}tates with {S}ub-\uppercase{P}lanck-scale {S}tructures},\ }\href
  {https://doi.org/10.1103/PhysRevLett.89.154103} {\bibfield  {journal}
  {\bibinfo  {journal} {Phys. Rev. Lett.}\ }\textbf {\bibinfo {volume} {89}},\
  \bibinfo {pages} {154103} (\bibinfo {year} {2002})}\BibitemShut {NoStop}%
\bibitem [{\citenamefont {Wisniacki}(2003)}]{Eff2}%
  \BibitemOpen
  \bibfield  {author} {\bibinfo {author} {\bibfnamefont {D.~A.}\ \bibnamefont
  {Wisniacki}},\ }\bibfield  {title} {\bibinfo {title} {Short-time decay of the
  {L}oschmidt echo},\ }\href {https://doi.org/10.1103/PhysRevE.67.016205}
  {\bibfield  {journal} {\bibinfo  {journal} {Phys. Rev. E}\ }\textbf {\bibinfo
  {volume} {67}},\ \bibinfo {pages} {016205} (\bibinfo {year}
  {2003})}\BibitemShut {NoStop}%
\bibitem [{\citenamefont {Ghosh}\ \emph {et~al.}(2006)\citenamefont {Ghosh},
  \citenamefont {Chiruvelli}, \citenamefont {Banerji},\ and\ \citenamefont
  {Panigrahi}}]{Eff3}%
  \BibitemOpen
  \bibfield  {author} {\bibinfo {author} {\bibfnamefont {S.}~\bibnamefont
  {Ghosh}}, \bibinfo {author} {\bibfnamefont {A.}~\bibnamefont {Chiruvelli}},
  \bibinfo {author} {\bibfnamefont {J.}~\bibnamefont {Banerji}},\ and\ \bibinfo
  {author} {\bibfnamefont {P.~K.}\ \bibnamefont {Panigrahi}},\ }\bibfield
  {title} {\bibinfo {title} {Mesoscopic superposition and
  sub-\uppercase{P}lanck-scale structure in molecular wave packets},\ }\href
  {https://doi.org/10.1103/PhysRevA.73.013411} {\bibfield  {journal} {\bibinfo
  {journal} {Phys. Rev. A}\ }\textbf {\bibinfo {volume} {73}},\ \bibinfo
  {pages} {013411} (\bibinfo {year} {2006})}\BibitemShut {NoStop}%
\bibitem [{\citenamefont {Praxmeyer}\ \emph {et~al.}(2007)\citenamefont
  {Praxmeyer}, \citenamefont {Wasylczyk}, \citenamefont {Radzewicz},\ and\
  \citenamefont {W\'odkiewicz}}]{Eff5}%
  \BibitemOpen
  \bibfield  {author} {\bibinfo {author} {\bibfnamefont {L.}~\bibnamefont
  {Praxmeyer}}, \bibinfo {author} {\bibfnamefont {P.}~\bibnamefont
  {Wasylczyk}}, \bibinfo {author} {\bibfnamefont {C.}~\bibnamefont
  {Radzewicz}},\ and\ \bibinfo {author} {\bibfnamefont {K.}~\bibnamefont
  {W\'odkiewicz}},\ }\bibfield  {title} {\bibinfo {title} {Time-{F}requency
  {D}omain {A}nalogues of {P}hase {S}pace {S}ub-\uppercase{P}lanck
  {S}tructures},\ }\href {https://doi.org/10.1103/PhysRevLett.98.063901}
  {\bibfield  {journal} {\bibinfo  {journal} {Phys. Rev. Lett.}\ }\textbf
  {\bibinfo {volume} {98}},\ \bibinfo {pages} {063901} (\bibinfo {year}
  {2007})}\BibitemShut {NoStop}%
\bibitem [{\citenamefont {Scott}\ and\ \citenamefont {Caves}(2008)}]{Eff6}%
  \BibitemOpen
  \bibfield  {author} {\bibinfo {author} {\bibfnamefont {A.~J.}\ \bibnamefont
  {Scott}}\ and\ \bibinfo {author} {\bibfnamefont {C.~M.}\ \bibnamefont
  {Caves}},\ }\bibfield  {title} {\bibinfo {title} {Teleportation fidelity as a
  probe of sub-\uppercase{P}lanck phase-space structure},\ }\href
  {https://doi.org/https://doi.org/10.1016/j.aop.2008.01.007} {\bibfield
  {journal} {\bibinfo  {journal} {Ann. Phys. (NY)}\ }\textbf {\bibinfo {volume}
  {323}},\ \bibinfo {pages} {2685} (\bibinfo {year} {2008})}\BibitemShut
  {NoStop}%
\bibitem [{\citenamefont {Bhatt}\ \emph {et~al.}(2008)\citenamefont {Bhatt},
  \citenamefont {Panigrahi},\ and\ \citenamefont {Vyas}}]{Eff7}%
  \BibitemOpen
  \bibfield  {author} {\bibinfo {author} {\bibfnamefont {J.~R.}\ \bibnamefont
  {Bhatt}}, \bibinfo {author} {\bibfnamefont {P.~K.}\ \bibnamefont
  {Panigrahi}},\ and\ \bibinfo {author} {\bibfnamefont {M.}~\bibnamefont
  {Vyas}},\ }\bibfield  {title} {\bibinfo {title} {Entanglement-induced
  sub-\uppercase{P}lanck phase-space structures},\ }\href
  {https://doi.org/10.1103/PhysRevA.78.034101} {\bibfield  {journal} {\bibinfo
  {journal} {Phys. Rev. A}\ }\textbf {\bibinfo {volume} {78}},\ \bibinfo
  {pages} {034101} (\bibinfo {year} {2008})}\BibitemShut {NoStop}%
\bibitem [{\citenamefont {Stobi\'{n}ska}\ \emph {et~al.}(2008)\citenamefont
  {Stobi\'{n}ska}, \citenamefont {Milburn},\ and\ \citenamefont
  {W\'odkiewicz}}]{Eff8}%
  \BibitemOpen
  \bibfield  {author} {\bibinfo {author} {\bibfnamefont {M.}~\bibnamefont
  {Stobi\'{n}ska}}, \bibinfo {author} {\bibfnamefont {G.~J.}\ \bibnamefont
  {Milburn}},\ and\ \bibinfo {author} {\bibfnamefont {K.}~\bibnamefont
  {W\'odkiewicz}},\ }\bibfield  {title} {\bibinfo {title} {\uppercase{W}igner
  function evolution of quantum states in the presence of self-{K}err
  interaction},\ }\href {https://doi.org/10.1103/PhysRevA.78.013810} {\bibfield
   {journal} {\bibinfo  {journal} {Phys. Rev. A}\ }\textbf {\bibinfo {volume}
  {78}},\ \bibinfo {pages} {013810} (\bibinfo {year} {2008})}\BibitemShut
  {NoStop}%
\bibitem [{\citenamefont {Ghosh}\ \emph {et~al.}(2009)\citenamefont {Ghosh},
  \citenamefont {Roy}, \citenamefont {Genes},\ and\ \citenamefont
  {Vitali}}]{Eff9}%
  \BibitemOpen
  \bibfield  {author} {\bibinfo {author} {\bibfnamefont {S.}~\bibnamefont
  {Ghosh}}, \bibinfo {author} {\bibfnamefont {U.}~\bibnamefont {Roy}}, \bibinfo
  {author} {\bibfnamefont {C.}~\bibnamefont {Genes}},\ and\ \bibinfo {author}
  {\bibfnamefont {D.}~\bibnamefont {Vitali}},\ }\bibfield  {title} {\bibinfo
  {title} {Sub-\uppercase{P}lanck-scale structures in a vibrating molecule in
  the presence of decoherence},\ }\href
  {https://doi.org/10.1103/PhysRevA.79.052104} {\bibfield  {journal} {\bibinfo
  {journal} {Phys. Rev. A}\ }\textbf {\bibinfo {volume} {79}},\ \bibinfo
  {pages} {052104} (\bibinfo {year} {2009})}\BibitemShut {NoStop}%
\bibitem [{\citenamefont {Roy}\ \emph {et~al.}(2009)\citenamefont {Roy},
  \citenamefont {Ghosh}, \citenamefont {Panigrahi},\ and\ \citenamefont
  {Vitali}}]{Eff10}%
  \BibitemOpen
  \bibfield  {author} {\bibinfo {author} {\bibfnamefont {U.}~\bibnamefont
  {Roy}}, \bibinfo {author} {\bibfnamefont {S.}~\bibnamefont {Ghosh}}, \bibinfo
  {author} {\bibfnamefont {P.~K.}\ \bibnamefont {Panigrahi}},\ and\ \bibinfo
  {author} {\bibfnamefont {D.}~\bibnamefont {Vitali}},\ }\bibfield  {title}
  {\bibinfo {title} {Sub-\uppercase{P}lanck-scale structures in the
  \uppercase{P}\"oschl-teller potential and their sensitivity to
  perturbations},\ }\href {https://doi.org/10.1103/PhysRevA.80.052115}
  {\bibfield  {journal} {\bibinfo  {journal} {Phys. Rev. A}\ }\textbf {\bibinfo
  {volume} {80}},\ \bibinfo {pages} {052115} (\bibinfo {year}
  {2009})}\BibitemShut {NoStop}%
\bibitem [{\citenamefont {Ghosh}(2012)}]{Eff11}%
  \BibitemOpen
  \bibfield  {author} {\bibinfo {author} {\bibfnamefont {S.}~\bibnamefont
  {Ghosh}},\ }\bibfield  {title} {\bibinfo {title} {Coherent control of
  mesoscopic superpositions in a diatomic molecule},\ }\href
  {https://doi.org/10.1142/S0219749912500141} {\bibfield  {journal} {\bibinfo
  {journal} {Int. J. Quantum Inf.}\ }\textbf {\bibinfo {volume} {10}},\
  \bibinfo {pages} {1250014} (\bibinfo {year} {2012})}\BibitemShut {NoStop}%
\bibitem [{\citenamefont {Kumari}\ \emph {et~al.}(2015)\citenamefont {Kumari},
  \citenamefont {Pan},\ and\ \citenamefont {Panigrahi}}]{Eff12}%
  \BibitemOpen
  \bibfield  {author} {\bibinfo {author} {\bibfnamefont {A.}~\bibnamefont
  {Kumari}}, \bibinfo {author} {\bibfnamefont {A.~K.}\ \bibnamefont {Pan}},\
  and\ \bibinfo {author} {\bibfnamefont {P.~K.}\ \bibnamefont {Panigrahi}},\
  }\bibfield  {title} {\bibinfo {title} {Sub-\uppercase{P}lanck structure in a
  mixed state},\ }\href {https://doi.org/10.1140/epjd/e2015-60269-2} {\bibfield
   {journal} {\bibinfo  {journal} {Eur. Phys. D}\ }\textbf {\bibinfo {volume}
  {69}},\ \bibinfo {pages} {248} (\bibinfo {year} {2015})}\BibitemShut
  {NoStop}%
\bibitem [{\citenamefont {Dodonov}\ \emph {et~al.}(2016)\citenamefont
  {Dodonov}, \citenamefont {Valverde}, \citenamefont {Souza},\ and\
  \citenamefont {Baseia}}]{Eff13}%
  \BibitemOpen
  \bibfield  {author} {\bibinfo {author} {\bibfnamefont {V.}~\bibnamefont
  {Dodonov}}, \bibinfo {author} {\bibfnamefont {C.}~\bibnamefont {Valverde}},
  \bibinfo {author} {\bibfnamefont {L.}~\bibnamefont {Souza}},\ and\ \bibinfo
  {author} {\bibfnamefont {B.}~\bibnamefont {Baseia}},\ }\bibfield  {title}
  {\bibinfo {title} {Decoherence of odd compass states in the phase-sensitive
  amplifying/dissipating environment},\ }\href
  {https://doi.org/https://doi.org/10.1016/j.aop.2016.04.019} {\bibfield
  {journal} {\bibinfo  {journal} {Ann. Phys. (NY)}\ }\textbf {\bibinfo {volume}
  {371}},\ \bibinfo {pages} {296} (\bibinfo {year} {2016})}\BibitemShut
  {NoStop}%
\bibitem [{\citenamefont {Kumar}\ and\ \citenamefont {Lee}(2017)}]{Eff14}%
  \BibitemOpen
  \bibfield  {author} {\bibinfo {author} {\bibfnamefont {P.}~\bibnamefont
  {Kumar}}\ and\ \bibinfo {author} {\bibfnamefont {R.-K.}\ \bibnamefont
  {Lee}},\ }\bibfield  {title} {\bibinfo {title} {Sensitivity of
  sub-\uppercase{P}lanck structures of mesoscopically superposed coherent
  states to the thermal reservoirs induced decoherence},\ }\href
  {https://doi.org/https://doi.org/10.1016/j.optcom.2017.02.066} {\bibfield
  {journal} {\bibinfo  {journal} {Opt. Commun.}\ }\textbf {\bibinfo {volume}
  {394}},\ \bibinfo {pages} {23} (\bibinfo {year} {2017})}\BibitemShut
  {NoStop}%
\bibitem [{\citenamefont {Howard}\ \emph {et~al.}(2019)\citenamefont {Howard},
  \citenamefont {Weinhold}, \citenamefont {Shahandeh}, \citenamefont {Combes},
  \citenamefont {Vanner}, \citenamefont {White},\ and\ \citenamefont
  {Ringbauer}}]{Eff15}%
  \BibitemOpen
  \bibfield  {author} {\bibinfo {author} {\bibfnamefont {L.~A.}\ \bibnamefont
  {Howard}}, \bibinfo {author} {\bibfnamefont {T.~J.}\ \bibnamefont
  {Weinhold}}, \bibinfo {author} {\bibfnamefont {F.}~\bibnamefont {Shahandeh}},
  \bibinfo {author} {\bibfnamefont {J.}~\bibnamefont {Combes}}, \bibinfo
  {author} {\bibfnamefont {M.~R.}\ \bibnamefont {Vanner}}, \bibinfo {author}
  {\bibfnamefont {A.~G.}\ \bibnamefont {White}},\ and\ \bibinfo {author}
  {\bibfnamefont {M.}~\bibnamefont {Ringbauer}},\ }\bibfield  {title} {\bibinfo
  {title} {Quantum {H}ypercube {S}tates},\ }\href
  {https://doi.org/10.1103/PhysRevLett.123.020402} {\bibfield  {journal}
  {\bibinfo  {journal} {Phys. Rev. Lett.}\ }\textbf {\bibinfo {volume} {123}},\
  \bibinfo {pages} {020402} (\bibinfo {year} {2019})}\BibitemShut {NoStop}%
\bibitem [{\citenamefont {Agarwal}\ and\ \citenamefont {Pathak}(2004)}]{Prop1}%
  \BibitemOpen
  \bibfield  {author} {\bibinfo {author} {\bibfnamefont {G.~S.}\ \bibnamefont
  {Agarwal}}\ and\ \bibinfo {author} {\bibfnamefont {P.~K.}\ \bibnamefont
  {Pathak}},\ }\bibfield  {title} {\bibinfo {title} {Mesoscopic superposition
  of states with sub-\uppercase{P}lanck structures in phase space},\ }\href
  {https://doi.org/10.1103/PhysRevA.70.053813} {\bibfield  {journal} {\bibinfo
  {journal} {Phys. Rev. A}\ }\textbf {\bibinfo {volume} {70}},\ \bibinfo
  {pages} {053813} (\bibinfo {year} {2004})}\BibitemShut {NoStop}%
\bibitem [{\citenamefont {Pathak}\ and\ \citenamefont {Agarwal}(2005)}]{Prop2}%
  \BibitemOpen
  \bibfield  {author} {\bibinfo {author} {\bibfnamefont {P.~K.}\ \bibnamefont
  {Pathak}}\ and\ \bibinfo {author} {\bibfnamefont {G.~S.}\ \bibnamefont
  {Agarwal}},\ }\bibfield  {title} {\bibinfo {title} {Generation of a
  superposition of multiple mesoscopic states of radiation in a resonant
  cavity},\ }\href {https://doi.org/10.1103/PhysRevA.71.043823} {\bibfield
  {journal} {\bibinfo  {journal} {Phys. Rev. A}\ }\textbf {\bibinfo {volume}
  {71}},\ \bibinfo {pages} {043823} (\bibinfo {year} {2005})}\BibitemShut
  {NoStop}%
\bibitem [{\citenamefont {Stobi{\'{n}}ska}\ \emph {et~al.}(2011)\citenamefont
  {Stobi{\'{n}}ska}, \citenamefont {Villar},\ and\ \citenamefont
  {Leuchs}}]{Prop3}%
  \BibitemOpen
  \bibfield  {author} {\bibinfo {author} {\bibfnamefont {M.}~\bibnamefont
  {Stobi{\'{n}}ska}}, \bibinfo {author} {\bibfnamefont {A.~S.}\ \bibnamefont
  {Villar}},\ and\ \bibinfo {author} {\bibfnamefont {G.}~\bibnamefont
  {Leuchs}},\ }\bibfield  {title} {\bibinfo {title} {Generation of
  \uppercase{K}err non-\uppercase{g}aussian motional states of trapped ions},\
  }\href {https://doi.org/10.1209/0295-5075/94/54002} {\bibfield  {journal}
  {\bibinfo  {journal} {Europhys. Lett.}\ }\textbf {\bibinfo {volume} {94}},\
  \bibinfo {pages} {54002} (\bibinfo {year} {2011})}\BibitemShut {NoStop}%
\bibitem [{\citenamefont {Govia}\ \emph {et~al.}(2014)\citenamefont {Govia},
  \citenamefont {Pritchett},\ and\ \citenamefont {Wilhelm}}]{Prop5}%
  \BibitemOpen
  \bibfield  {author} {\bibinfo {author} {\bibfnamefont {L.~C.~G.}\
  \bibnamefont {Govia}}, \bibinfo {author} {\bibfnamefont {E.~J.}\ \bibnamefont
  {Pritchett}},\ and\ \bibinfo {author} {\bibfnamefont {F.~K.}\ \bibnamefont
  {Wilhelm}},\ }\bibfield  {title} {\bibinfo {title} {Generating nonclassical
  states from classical radiation by subtraction measurements},\ }\href
  {https://doi.org/10.1088/1367-2630/16/4/045011} {\bibfield  {journal}
  {\bibinfo  {journal} {New J. Phys.}\ }\textbf {\bibinfo {volume} {16}},\
  \bibinfo {pages} {045011} (\bibinfo {year} {2014})}\BibitemShut {NoStop}%
\bibitem [{\citenamefont {Hastrup}\ \emph {et~al.}(2020)\citenamefont
  {Hastrup}, \citenamefont {Neergaard-Nielsen},\ and\ \citenamefont
  {Andersen}}]{Prop6}%
  \BibitemOpen
  \bibfield  {author} {\bibinfo {author} {\bibfnamefont {J.}~\bibnamefont
  {Hastrup}}, \bibinfo {author} {\bibfnamefont {J.~S.}\ \bibnamefont
  {Neergaard-Nielsen}},\ and\ \bibinfo {author} {\bibfnamefont {U.~L.}\
  \bibnamefont {Andersen}},\ }\bibfield  {title} {\bibinfo {title}
  {Deterministic generation of a four-component optical cat state},\ }\href
  {https://doi.org/10.1364/OL.383194} {\bibfield  {journal} {\bibinfo
  {journal} {Opt. Lett.}\ }\textbf {\bibinfo {volume} {45}},\ \bibinfo {pages}
  {640} (\bibinfo {year} {2020})}\BibitemShut {NoStop}%
\bibitem [{\citenamefont {Ofek}\ \emph {et~al.}(2016)\citenamefont {Ofek},
  \citenamefont {Petrenko}, \citenamefont {Heeres}, \citenamefont {Reinhold},
  \citenamefont {Leghtas}, \citenamefont {Vlastakis}, \citenamefont {Liu},
  \citenamefont {Frunzio}, \citenamefont {Girvin}, \citenamefont {Jiang},
  \citenamefont {Mirrahimi}, \citenamefont {Devoret},\ and\ \citenamefont
  {Schoelkopf}}]{Exp1}%
  \BibitemOpen
  \bibfield  {author} {\bibinfo {author} {\bibfnamefont {N.}~\bibnamefont
  {Ofek}}, \bibinfo {author} {\bibfnamefont {A.}~\bibnamefont {Petrenko}},
  \bibinfo {author} {\bibfnamefont {R.}~\bibnamefont {Heeres}}, \bibinfo
  {author} {\bibfnamefont {P.}~\bibnamefont {Reinhold}}, \bibinfo {author}
  {\bibfnamefont {Z.}~\bibnamefont {Leghtas}}, \bibinfo {author} {\bibfnamefont
  {B.}~\bibnamefont {Vlastakis}}, \bibinfo {author} {\bibfnamefont
  {Y.}~\bibnamefont {Liu}}, \bibinfo {author} {\bibfnamefont {L.}~\bibnamefont
  {Frunzio}}, \bibinfo {author} {\bibfnamefont {S.~M.}\ \bibnamefont {Girvin}},
  \bibinfo {author} {\bibfnamefont {L.}~\bibnamefont {Jiang}}, \bibinfo
  {author} {\bibfnamefont {M.}~\bibnamefont {Mirrahimi}}, \bibinfo {author}
  {\bibfnamefont {M.~H.}\ \bibnamefont {Devoret}},\ and\ \bibinfo {author}
  {\bibfnamefont {R.~J.}\ \bibnamefont {Schoelkopf}},\ }\bibfield  {title}
  {\bibinfo {title} {Extending the lifetime of a quantum bit with error
  correction in superconducting circuits},\ }\href
  {https://doi.org/10.1038/nature18949} {\bibfield  {journal} {\bibinfo
  {journal} {Nature}\ }\textbf {\bibinfo {volume} {536}},\ \bibinfo {pages}
  {441} (\bibinfo {year} {2016})}\BibitemShut {NoStop}%
\bibitem [{\citenamefont {Vlastakis}\ \emph {et~al.}(2013)\citenamefont
  {Vlastakis}, \citenamefont {Kirchmair}, \citenamefont {Leghtas},
  \citenamefont {Nigg}, \citenamefont {Frunzio}, \citenamefont {Girvin},
  \citenamefont {Mirrahimi}, \citenamefont {Devoret},\ and\ \citenamefont
  {Schoelkopf}}]{Exp2}%
  \BibitemOpen
  \bibfield  {author} {\bibinfo {author} {\bibfnamefont {B.}~\bibnamefont
  {Vlastakis}}, \bibinfo {author} {\bibfnamefont {G.}~\bibnamefont
  {Kirchmair}}, \bibinfo {author} {\bibfnamefont {Z.}~\bibnamefont {Leghtas}},
  \bibinfo {author} {\bibfnamefont {S.~E.}\ \bibnamefont {Nigg}}, \bibinfo
  {author} {\bibfnamefont {L.}~\bibnamefont {Frunzio}}, \bibinfo {author}
  {\bibfnamefont {S.~M.}\ \bibnamefont {Girvin}}, \bibinfo {author}
  {\bibfnamefont {M.}~\bibnamefont {Mirrahimi}}, \bibinfo {author}
  {\bibfnamefont {M.~H.}\ \bibnamefont {Devoret}},\ and\ \bibinfo {author}
  {\bibfnamefont {R.~J.}\ \bibnamefont {Schoelkopf}},\ }\bibfield  {title}
  {\bibinfo {title} {Deterministically encoding quantum information using
  100-photon \uppercase{S}chr{\"o}dinger cat states},\ }\href
  {https://doi.org/10.1126/science.1243289} {\bibfield  {journal} {\bibinfo
  {journal} {Science}\ }\textbf {\bibinfo {volume} {342}},\ \bibinfo {pages}
  {607} (\bibinfo {year} {2013})}\BibitemShut {NoStop}%
\bibitem [{\citenamefont {Praxmeyer}\ \emph {et~al.}(2016)\citenamefont
  {Praxmeyer}, \citenamefont {Chen}, \citenamefont {Yang}, \citenamefont
  {Yang},\ and\ \citenamefont {Lee}}]{Exp3}%
  \BibitemOpen
  \bibfield  {author} {\bibinfo {author} {\bibfnamefont {L.}~\bibnamefont
  {Praxmeyer}}, \bibinfo {author} {\bibfnamefont {C.-C.}\ \bibnamefont {Chen}},
  \bibinfo {author} {\bibfnamefont {P.}~\bibnamefont {Yang}}, \bibinfo {author}
  {\bibfnamefont {S.~D.}\ \bibnamefont {Yang}},\ and\ \bibinfo {author}
  {\bibfnamefont {R.~K.}\ \bibnamefont {Lee}},\ }\bibfield  {title} {\bibinfo
  {title} {Direct measurement of time-frequency analogs of
  sub-\uppercase{P}lanck structures},\ }\href
  {https://doi.org/10.1103/PhysRevA.93.053835} {\bibfield  {journal} {\bibinfo
  {journal} {Phys. Rev. A}\ }\textbf {\bibinfo {volume} {93}},\ \bibinfo
  {pages} {053835} (\bibinfo {year} {2016})}\BibitemShut {NoStop}%
\bibitem [{\citenamefont {Lemos}\ \emph {et~al.}(2012)\citenamefont {Lemos},
  \citenamefont {Gomes}, \citenamefont {Walborn}, \citenamefont {S.~Ribeiro},\
  and\ \citenamefont {Toscano}}]{Exp4}%
  \BibitemOpen
  \bibfield  {author} {\bibinfo {author} {\bibfnamefont {G.~B.}\ \bibnamefont
  {Lemos}}, \bibinfo {author} {\bibfnamefont {R.~M.}\ \bibnamefont {Gomes}},
  \bibinfo {author} {\bibfnamefont {S.~P.}\ \bibnamefont {Walborn}}, \bibinfo
  {author} {\bibfnamefont {P.~H.}\ \bibnamefont {S.~Ribeiro}},\ and\ \bibinfo
  {author} {\bibfnamefont {F.}~\bibnamefont {Toscano}},\ }\bibfield  {title}
  {\bibinfo {title} {Experimental observation of quantum chaos in a beam of
  light},\ }\href {https://doi.org/10.1038/ncomms2214} {\bibfield  {journal}
  {\bibinfo  {journal} {Nat. Commun.}\ }\textbf {\bibinfo {volume} {3}},\
  \bibinfo {pages} {2041} (\bibinfo {year} {2012})}\BibitemShut {NoStop}%
\bibitem [{\citenamefont {Austin}\ \emph {et~al.}(2010)\citenamefont {Austin},
  \citenamefont {Witting}, \citenamefont {Wyatt},\ and\ \citenamefont
  {Walmsley}}]{Exp5}%
  \BibitemOpen
  \bibfield  {author} {\bibinfo {author} {\bibfnamefont {D.~R.}\ \bibnamefont
  {Austin}}, \bibinfo {author} {\bibfnamefont {T.}~\bibnamefont {Witting}},
  \bibinfo {author} {\bibfnamefont {A.~S.}\ \bibnamefont {Wyatt}},\ and\
  \bibinfo {author} {\bibfnamefont {I.~A.}\ \bibnamefont {Walmsley}},\
  }\bibfield  {title} {\bibinfo {title} {Measuring sub-\uppercase{P}lanck
  structural analogues in chronocyclic phase space},\ }\href
  {https://doi.org/https://doi.org/10.1016/j.optcom.2009.10.060} {\bibfield
  {journal} {\bibinfo  {journal} {Opt. Commun.}\ }\textbf {\bibinfo {volume}
  {283}},\ \bibinfo {pages} {855} (\bibinfo {year} {2010})}\BibitemShut
  {NoStop}%
\bibitem [{\citenamefont {Weyl}(1950{\natexlab{b}})}]{weyl1950theory}%
  \BibitemOpen
  \bibfield  {author} {\bibinfo {author} {\bibfnamefont {H.}~\bibnamefont
  {Weyl}},\ }\href@noop {} {\emph {\bibinfo {title} {The Theory of Groups and
  Quantum Mechanics}}}\ (\bibinfo  {publisher} {Dover},\ \bibinfo {address}
  {New York},\ \bibinfo {year} {1950})\BibitemShut {NoStop}%
\bibitem [{\citenamefont {Schr\"{o}dinger}(1926)}]{Sch26}%
  \BibitemOpen
  \bibfield  {author} {\bibinfo {author} {\bibfnamefont {E.}~\bibnamefont
  {Schr\"{o}dinger}},\ }\bibfield  {title} {\bibinfo {title} {Der stetige
  \"{u}bergang von der mikro- zur makromechanik},\ }\href
  {https://doi.org/https://doi.org/10.1007/BF01507634} {\bibfield  {journal}
  {\bibinfo  {journal} {Sci. Nat.}\ }\textbf {\bibinfo {volume} {14}},\
  \bibinfo {pages} {664} (\bibinfo {year} {1926})}\BibitemShut {NoStop}%
\bibitem [{\citenamefont {Milonni}\ and\ \citenamefont {Nieto}(2009)}]{MN09}%
  \BibitemOpen
  \bibfield  {author} {\bibinfo {author} {\bibfnamefont {P.~W.}\ \bibnamefont
  {Milonni}}\ and\ \bibinfo {author} {\bibfnamefont {M.~M.}\ \bibnamefont
  {Nieto}},\ }\bibfield  {title} {\bibinfo {title} {{C}oherent states},\ }in\
  \href@noop {} {\emph {\bibinfo {booktitle} {{C}ompendium of {Q}uantum
  {P}hysics,}}}\ (\bibinfo  {publisher} {Springer},\ \bibinfo {address}
  {Berlin, Heidelberg},\ \bibinfo {year} {2009})\BibitemShut {NoStop}%
\bibitem [{\citenamefont {Glauber}(1963)}]{Gla63}%
  \BibitemOpen
  \bibfield  {author} {\bibinfo {author} {\bibfnamefont {R.~J.}\ \bibnamefont
  {Glauber}},\ }\bibfield  {title} {\bibinfo {title} {Coherent and incoherent
  states of the radiation field},\ }\href
  {https://doi.org/10.1103/PhysRev.131.2766} {\bibfield  {journal} {\bibinfo
  {journal} {Phys. Rev.}\ }\textbf {\bibinfo {volume} {131}},\ \bibinfo {pages}
  {2766} (\bibinfo {year} {1963})}\BibitemShut {NoStop}%
\bibitem [{\citenamefont {Perelomov}(1986)}]{Per86}%
  \BibitemOpen
  \bibfield  {author} {\bibinfo {author} {\bibfnamefont {A.}~\bibnamefont
  {Perelomov}},\ }\href@noop {} {\emph {\bibinfo {title} {Generalized Coherent
  States and Their Applications}}},\ Theoretical and Mathematical Physics\
  (\bibinfo  {publisher} {Springer-Verlag},\ \bibinfo {address} {Berlin},\
  \bibinfo {year} {1986})\BibitemShut {NoStop}%
\bibitem [{\citenamefont {Gazeau}(2009)}]{Gaz09}%
  \BibitemOpen
  \bibfield  {author} {\bibinfo {author} {\bibfnamefont {J.-P.}\ \bibnamefont
  {Gazeau}},\ }\href@noop {} {\emph {\bibinfo {title} {Coherent States in
  Quantum Physics}}}\ (\bibinfo  {publisher} {Wiley-VCH},\ \bibinfo {address}
  {Berlin},\ \bibinfo {year} {2009})\BibitemShut {NoStop}%
\bibitem [{\citenamefont {Shukla}\ \emph {et~al.}(2019)\citenamefont {Shukla},
  \citenamefont {Akhtar},\ and\ \citenamefont {Sanders}}]{PhysRevA.99.063813}%
  \BibitemOpen
  \bibfield  {author} {\bibinfo {author} {\bibfnamefont {N.}~\bibnamefont
  {Shukla}}, \bibinfo {author} {\bibfnamefont {N.}~\bibnamefont {Akhtar}},\
  and\ \bibinfo {author} {\bibfnamefont {B.~C.}\ \bibnamefont {Sanders}},\
  }\bibfield  {title} {\bibinfo {title} {Quantum tetrachotomous states:
  Superposition of four coherent states on a line in phase space},\ }\href
  {https://doi.org/10.1103/PhysRevA.99.063813} {\bibfield  {journal} {\bibinfo
  {journal} {Phys. Rev. A}\ }\textbf {\bibinfo {volume} {99}},\ \bibinfo
  {pages} {063813} (\bibinfo {year} {2019})}\BibitemShut {NoStop}%
\bibitem [{\citenamefont {Milburn}(1986)}]{Mil86}%
  \BibitemOpen
  \bibfield  {author} {\bibinfo {author} {\bibfnamefont {G.~J.}\ \bibnamefont
  {Milburn}},\ }\bibfield  {title} {\bibinfo {title} {Quantum and classical
  {L}iouville dynamics of the anharmonic oscillator},\ }\href
  {https://doi.org/10.1103/PhysRevA.33.674} {\bibfield  {journal} {\bibinfo
  {journal} {Phys. Rev. A}\ }\textbf {\bibinfo {volume} {33}},\ \bibinfo
  {pages} {674} (\bibinfo {year} {1986})}\BibitemShut {NoStop}%
\bibitem [{\citenamefont {Yurke}\ and\ \citenamefont {Stoler}(1986)}]{YS86}%
  \BibitemOpen
  \bibfield  {author} {\bibinfo {author} {\bibfnamefont {B.}~\bibnamefont
  {Yurke}}\ and\ \bibinfo {author} {\bibfnamefont {D.}~\bibnamefont {Stoler}},\
  }\bibfield  {title} {\bibinfo {title} {Generating {Q}uantum {M}echanical
  {S}uperpositions of {M}acroscopically {D}istinguishable {S}tates via
  {A}mplitude {D}ispersion},\ }\href
  {https://doi.org/10.1103/PhysRevLett.57.13} {\bibfield  {journal} {\bibinfo
  {journal} {Phys. Rev. Lett.}\ }\textbf {\bibinfo {volume} {57}},\ \bibinfo
  {pages} {13} (\bibinfo {year} {1986})}\BibitemShut {NoStop}%
\bibitem [{\citenamefont {Akhtar}\ \emph {et~al.}(2021)\citenamefont {Akhtar},
  \citenamefont {Sanders},\ and\ \citenamefont
  {Navarrete-Benlloch}}]{Naeem2021}%
  \BibitemOpen
  \bibfield  {author} {\bibinfo {author} {\bibfnamefont {N.}~\bibnamefont
  {Akhtar}}, \bibinfo {author} {\bibfnamefont {B.~C.}\ \bibnamefont
  {Sanders}},\ and\ \bibinfo {author} {\bibfnamefont {C.}~\bibnamefont
  {Navarrete-Benlloch}},\ }\bibfield  {title} {\bibinfo {title} {Sub-planck
  structures: Analogies between the {H}eisenberg-{W}eyl and {SU}(2) groups},\
  }\href {https://doi.org/10.1103/PhysRevA.103.053711} {\bibfield  {journal}
  {\bibinfo  {journal} {Phys. Rev. A}\ }\textbf {\bibinfo {volume} {103}},\
  \bibinfo {pages} {053711} (\bibinfo {year} {2021})}\BibitemShut {NoStop}%
\bibitem [{\citenamefont {Gerry}(1985)}]{Gerry1986}%
  \BibitemOpen
  \bibfield  {author} {\bibinfo {author} {\bibfnamefont {C.~C.}\ \bibnamefont
  {Gerry}},\ }\bibfield  {title} {\bibinfo {title} {Dynamics of {SU}(1,1)
  coherent states},\ }\href {https://doi.org/10.1103/PhysRevA.31.2721}
  {\bibfield  {journal} {\bibinfo  {journal} {Phys. Rev. A}\ }\textbf {\bibinfo
  {volume} {31}},\ \bibinfo {pages} {2721} (\bibinfo {year}
  {1985})}\BibitemShut {NoStop}%
\bibitem [{\citenamefont {Ojeda-Guillén}\ \emph {et~al.}(2014)\citenamefont
  {Ojeda-Guillén}, \citenamefont {Mota},\ and\ \citenamefont
  {Granados}}]{Ojeda2014}%
  \BibitemOpen
  \bibfield  {author} {\bibinfo {author} {\bibfnamefont {D.}~\bibnamefont
  {Ojeda-Guillén}}, \bibinfo {author} {\bibfnamefont {R.~D.}\ \bibnamefont
  {Mota}},\ and\ \bibinfo {author} {\bibfnamefont {V.~D.}\ \bibnamefont
  {Granados}},\ }\bibfield  {title} {\bibinfo {title} {The {SU}(1,1)
  {P}erelomov number coherent states and the non-degenerate parametric
  amplifier},\ }\href {https://doi.org/10.1063/1.4871445} {\bibfield  {journal}
  {\bibinfo  {journal} {J. Math. Phys.}\ }\textbf {\bibinfo {volume} {55}},\
  \bibinfo {pages} {042109} (\bibinfo {year} {2014})}\BibitemShut {NoStop}%
\bibitem [{\citenamefont {Wodkiewicz}\ and\ \citenamefont
  {Eberly}(1985)}]{Wodkiewicz1985}%
  \BibitemOpen
  \bibfield  {author} {\bibinfo {author} {\bibfnamefont {K.}~\bibnamefont
  {Wodkiewicz}}\ and\ \bibinfo {author} {\bibfnamefont {J.~H.}\ \bibnamefont
  {Eberly}},\ }\bibfield  {title} {\bibinfo {title} {Coherent states, squeezed
  fluctuations, and the {SU}(2) am {SU}(1,1) groups in quantum-optics
  applications},\ }\href {https://doi.org/10.1364/JOSAB.2.000458} {\bibfield
  {journal} {\bibinfo  {journal} {J. Opt. Soc. Am. B}\ }\textbf {\bibinfo
  {volume} {2}},\ \bibinfo {pages} {458} (\bibinfo {year} {1985})}\BibitemShut
  {NoStop}%
\bibitem [{\citenamefont {Chaturvedi}\ and\ \citenamefont
  {Srinivasan}(1991)}]{Chaturvedi1991}%
  \BibitemOpen
  \bibfield  {author} {\bibinfo {author} {\bibfnamefont {S.}~\bibnamefont
  {Chaturvedi}}\ and\ \bibinfo {author} {\bibfnamefont {V.}~\bibnamefont
  {Srinivasan}},\ }\bibfield  {title} {\bibinfo {title} {Class of exactly
  solvable master equations describing coupled nonlinear oscillators},\ }\href
  {https://doi.org/10.1103/PhysRevA.43.4054} {\bibfield  {journal} {\bibinfo
  {journal} {Phys. Rev. A}\ }\textbf {\bibinfo {volume} {43}},\ \bibinfo
  {pages} {4054} (\bibinfo {year} {1991})}\BibitemShut {NoStop}%
\bibitem [{\citenamefont {Ban}(1992)}]{Ban1992}%
  \BibitemOpen
  \bibfield  {author} {\bibinfo {author} {\bibfnamefont {M.}~\bibnamefont
  {Ban}},\ }\bibfield  {title} {\bibinfo {title} {{SU}(1,1) {L}ie algebraic
  approach to linear dissipative processes in quantum optics},\ }\href
  {https://doi.org/10.1063/1.529540} {\bibfield  {journal} {\bibinfo  {journal}
  {J. Math. Phys.}\ }\textbf {\bibinfo {volume} {33}},\ \bibinfo {pages} {3213}
  (\bibinfo {year} {1992})}\BibitemShut {NoStop}%
\bibitem [{\citenamefont {Ban}(1993{\natexlab{a}})}]{Ban1993}%
  \BibitemOpen
  \bibfield  {author} {\bibinfo {author} {\bibfnamefont {M.}~\bibnamefont
  {Ban}},\ }\bibfield  {title} {\bibinfo {title} {{L}ie-algebra methods in
  quantum optics: The {L}iouville-space formulation},\ }\href
  {https://doi.org/10.1103/PhysRevA.47.5093} {\bibfield  {journal} {\bibinfo
  {journal} {Phys. Rev. A}\ }\textbf {\bibinfo {volume} {47}},\ \bibinfo
  {pages} {5093} (\bibinfo {year} {1993}{\natexlab{a}})}\BibitemShut {NoStop}%
\bibitem [{\citenamefont {Ban}(1993{\natexlab{b}})}]{ban1993b}%
  \BibitemOpen
  \bibfield  {author} {\bibinfo {author} {\bibfnamefont {M.}~\bibnamefont
  {Ban}},\ }\bibfield  {title} {\bibinfo {title} {Decomposition formulas for
  su(1, 1) and su(2) {L}ie algebras and their applications in quantum optics},\
  }\href {https://doi.org/10.1364/JOSAB.10.001347} {\bibfield  {journal}
  {\bibinfo  {journal} {J. Opt. Soc. Am. B}\ }\textbf {\bibinfo {volume}
  {10}},\ \bibinfo {pages} {1347} (\bibinfo {year}
  {1993}{\natexlab{b}})}\BibitemShut {NoStop}%
\bibitem [{\citenamefont {Jing}\ \emph {et~al.}(2011)\citenamefont {Jing},
  \citenamefont {Liu}, \citenamefont {Zhou}, \citenamefont {Ou},\ and\
  \citenamefont {Zhang}}]{Jing2011}%
  \BibitemOpen
  \bibfield  {author} {\bibinfo {author} {\bibfnamefont {J.}~\bibnamefont
  {Jing}}, \bibinfo {author} {\bibfnamefont {C.}~\bibnamefont {Liu}}, \bibinfo
  {author} {\bibfnamefont {Z.}~\bibnamefont {Zhou}}, \bibinfo {author}
  {\bibfnamefont {Z.~Y.}\ \bibnamefont {Ou}},\ and\ \bibinfo {author}
  {\bibfnamefont {W.}~\bibnamefont {Zhang}},\ }\bibfield  {title} {\bibinfo
  {title} {Realization of a nonlinear interferometer with parametric
  amplifiers},\ }\href {https://doi.org/10.1063/1.3606549} {\bibfield
  {journal} {\bibinfo  {journal} {Appl. Phys. Lett.}\ }\textbf {\bibinfo
  {volume} {99}},\ \bibinfo {pages} {011110} (\bibinfo {year}
  {2011})}\BibitemShut {NoStop}%
\bibitem [{\citenamefont {Hudelist}\ \emph {et~al.}(2014)\citenamefont
  {Hudelist}, \citenamefont {Kong}, \citenamefont {Liu}, \citenamefont {Jing},
  \citenamefont {Ou},\ and\ \citenamefont {Zhang}}]{Hudelist2014}%
  \BibitemOpen
  \bibfield  {author} {\bibinfo {author} {\bibfnamefont {F.}~\bibnamefont
  {Hudelist}}, \bibinfo {author} {\bibfnamefont {J.}~\bibnamefont {Kong}},
  \bibinfo {author} {\bibfnamefont {C.}~\bibnamefont {Liu}}, \bibinfo {author}
  {\bibfnamefont {J.}~\bibnamefont {Jing}}, \bibinfo {author} {\bibfnamefont
  {Z.}~\bibnamefont {Ou}},\ and\ \bibinfo {author} {\bibfnamefont
  {W.}~\bibnamefont {Zhang}},\ }\bibfield  {title} {\bibinfo {title} {Quantum
  metrology with parametric amplifier-based photon correlation
  interferometers},\ }\href {https://doi.org/10.1038/ncomms4049} {\bibfield
  {journal} {\bibinfo  {journal} {Nature Commun.}\ }\textbf {\bibinfo {volume}
  {5}},\ \bibinfo {pages} {1} (\bibinfo {year} {2014})}\BibitemShut {NoStop}%
\bibitem [{\citenamefont {Berrada}(2013)}]{Berrada2013}%
  \BibitemOpen
  \bibfield  {author} {\bibinfo {author} {\bibfnamefont {K.}~\bibnamefont
  {Berrada}},\ }\bibfield  {title} {\bibinfo {title} {Quantum metrology with
  {SU}(1,1) coherent states in the presence of nonlinear phase shifts},\ }\href
  {https://doi.org/10.1103/PhysRevA.88.013817} {\bibfield  {journal} {\bibinfo
  {journal} {Phys. Rev. A}\ }\textbf {\bibinfo {volume} {88}},\ \bibinfo
  {pages} {013817} (\bibinfo {year} {2013})}\BibitemShut {NoStop}%
\bibitem [{\citenamefont {Yurke}\ \emph {et~al.}(1986)\citenamefont {Yurke},
  \citenamefont {McCall},\ and\ \citenamefont {Klauder}}]{Yurke1896}%
  \BibitemOpen
  \bibfield  {author} {\bibinfo {author} {\bibfnamefont {B.}~\bibnamefont
  {Yurke}}, \bibinfo {author} {\bibfnamefont {S.~L.}\ \bibnamefont {McCall}},\
  and\ \bibinfo {author} {\bibfnamefont {J.~R.}\ \bibnamefont {Klauder}},\
  }\bibfield  {title} {\bibinfo {title} {{SU}(2) and {SU}(1,1)
  interferometers},\ }\href {https://doi.org/10.1103/PhysRevA.33.4033}
  {\bibfield  {journal} {\bibinfo  {journal} {Phys. Rev. A}\ }\textbf {\bibinfo
  {volume} {33}},\ \bibinfo {pages} {4033} (\bibinfo {year}
  {1986})}\BibitemShut {NoStop}%
\bibitem [{\citenamefont {Li}\ \emph {et~al.}(2016)\citenamefont {Li},
  \citenamefont {Gard}, \citenamefont {Gao}, \citenamefont {Yuan},
  \citenamefont {Zhang}, \citenamefont {Lee},\ and\ \citenamefont
  {Dowling}}]{Dong2016}%
  \BibitemOpen
  \bibfield  {author} {\bibinfo {author} {\bibfnamefont {D.}~\bibnamefont
  {Li}}, \bibinfo {author} {\bibfnamefont {B.~T.}\ \bibnamefont {Gard}},
  \bibinfo {author} {\bibfnamefont {Y.}~\bibnamefont {Gao}}, \bibinfo {author}
  {\bibfnamefont {C.-H.}\ \bibnamefont {Yuan}}, \bibinfo {author}
  {\bibfnamefont {W.}~\bibnamefont {Zhang}}, \bibinfo {author} {\bibfnamefont
  {H.}~\bibnamefont {Lee}},\ and\ \bibinfo {author} {\bibfnamefont {J.~P.}\
  \bibnamefont {Dowling}},\ }\bibfield  {title} {\bibinfo {title} {Phase
  sensitivity at the {H}eisenberg limit in an {SU}(1,1) interferometer via
  parity detection},\ }\href {https://doi.org/10.1103/PhysRevA.94.063840}
  {\bibfield  {journal} {\bibinfo  {journal} {Phys. Rev. A}\ }\textbf {\bibinfo
  {volume} {94}},\ \bibinfo {pages} {063840} (\bibinfo {year}
  {2016})}\BibitemShut {NoStop}%
\bibitem [{\citenamefont {Szigeti}\ \emph {et~al.}(2017)\citenamefont
  {Szigeti}, \citenamefont {Lewis-Swan},\ and\ \citenamefont
  {Haine}}]{Szigeti2017}%
  \BibitemOpen
  \bibfield  {author} {\bibinfo {author} {\bibfnamefont {S.~S.}\ \bibnamefont
  {Szigeti}}, \bibinfo {author} {\bibfnamefont {R.~J.}\ \bibnamefont
  {Lewis-Swan}},\ and\ \bibinfo {author} {\bibfnamefont {S.~A.}\ \bibnamefont
  {Haine}},\ }\bibfield  {title} {\bibinfo {title} {Pumped-{U}p {SU}(1,1)
  {I}nterferometry},\ }\href {https://doi.org/10.1103/PhysRevLett.118.150401}
  {\bibfield  {journal} {\bibinfo  {journal} {Phys. Rev. Lett.}\ }\textbf
  {\bibinfo {volume} {118}},\ \bibinfo {pages} {150401} (\bibinfo {year}
  {2017})}\BibitemShut {NoStop}%
\bibitem [{\citenamefont {Liang}\ \emph {et~al.}(2020)\citenamefont {Liang},
  \citenamefont {Su}, \citenamefont {Xiao}, \citenamefont {Che}, \citenamefont
  {Sanders},\ and\ \citenamefont {Wang}}]{barry2020}%
  \BibitemOpen
  \bibfield  {author} {\bibinfo {author} {\bibfnamefont {H.}~\bibnamefont
  {Liang}}, \bibinfo {author} {\bibfnamefont {Y.}~\bibnamefont {Su}}, \bibinfo
  {author} {\bibfnamefont {X.}~\bibnamefont {Xiao}}, \bibinfo {author}
  {\bibfnamefont {Y.}~\bibnamefont {Che}}, \bibinfo {author} {\bibfnamefont
  {B.~C.}\ \bibnamefont {Sanders}},\ and\ \bibinfo {author} {\bibfnamefont
  {X.}~\bibnamefont {Wang}},\ }\bibfield  {title} {\bibinfo {title}
  {Criticality in two-mode interferometers},\ }\href
  {https://doi.org/10.1103/PhysRevA.102.013722} {\bibfield  {journal} {\bibinfo
   {journal} {Phys. Rev. A}\ }\textbf {\bibinfo {volume} {102}},\ \bibinfo
  {pages} {013722} (\bibinfo {year} {2020})}\BibitemShut {NoStop}%
\bibitem [{\citenamefont {Duan}(2022{\natexlab{a}})}]{duan2022a}%
  \BibitemOpen
  \bibfield  {author} {\bibinfo {author} {\bibfnamefont {L.}~\bibnamefont
  {Duan}},\ }\bibfield  {title} {\bibinfo {title} {Quantum walk for {SU}(1,
  1)},\ }\href@noop {} {\bibfield  {journal} {\bibinfo  {journal} {arXiv
  preprint arXiv:2207.04511}\ } (\bibinfo {year}
  {2022}{\natexlab{a}})}\BibitemShut {NoStop}%
\bibitem [{\citenamefont {Duan}(2022{\natexlab{b}})}]{duan2022b}%
  \BibitemOpen
  \bibfield  {author} {\bibinfo {author} {\bibfnamefont {L.}~\bibnamefont
  {Duan}},\ }\bibfield  {title} {\bibinfo {title} {Unified approach to the
  nonlinear {R}abi models},\ }\href {https://doi.org/10.1088/1367-2630/ac8a68}
  {\bibfield  {journal} {\bibinfo  {journal} {New J. Phys.}\ }\textbf {\bibinfo
  {volume} {24}},\ \bibinfo {pages} {083045} (\bibinfo {year}
  {2022}{\natexlab{b}})}\BibitemShut {NoStop}%
\bibitem [{\citenamefont {Lo}\ and\ \citenamefont {Liu}(1993)}]{Lo1993}%
  \BibitemOpen
  \bibfield  {author} {\bibinfo {author} {\bibfnamefont {C.~F.}\ \bibnamefont
  {Lo}}\ and\ \bibinfo {author} {\bibfnamefont {K.~L.}\ \bibnamefont {Liu}},\
  }\bibfield  {title} {\bibinfo {title} {Multimode bosonic realization of the
  {SU}(1,1) {L}ie algebra},\ }\href {https://doi.org/10.1103/PhysRevA.48.3362}
  {\bibfield  {journal} {\bibinfo  {journal} {Phys. Rev. A}\ }\textbf {\bibinfo
  {volume} {48}},\ \bibinfo {pages} {3362} (\bibinfo {year}
  {1993})}\BibitemShut {NoStop}%
\bibitem [{\citenamefont {Brif}\ \emph {et~al.}(1996)\citenamefont {Brif},
  \citenamefont {Vourdas},\ and\ \citenamefont {Mann}}]{Brif1996}%
  \BibitemOpen
  \bibfield  {author} {\bibinfo {author} {\bibfnamefont {C.}~\bibnamefont
  {Brif}}, \bibinfo {author} {\bibfnamefont {A.}~\bibnamefont {Vourdas}},\ and\
  \bibinfo {author} {\bibfnamefont {A.}~\bibnamefont {Mann}},\ }\bibfield
  {title} {\bibinfo {title} {Analytic representations based on {SU}(1,1)
  coherent states and their applications},\ }\href
  {https://doi.org/10.1088/0305-4470/29/18/017} {\bibfield  {journal} {\bibinfo
   {journal} {J. Phys. A: Math. Theor.}\ }\textbf {\bibinfo {volume} {29}},\
  \bibinfo {pages} {5873} (\bibinfo {year} {1996})}\BibitemShut {NoStop}%
\bibitem [{\citenamefont {Gerry}(1991{\natexlab{a}})}]{Gerry1991}%
  \BibitemOpen
  \bibfield  {author} {\bibinfo {author} {\bibfnamefont {C.~C.}\ \bibnamefont
  {Gerry}},\ }\bibfield  {title} {\bibinfo {title} {Correlated two-mode
  {SU}(1,1) coherent states: nonclassical properties},\ }\href
  {https://doi.org/10.1364/JOSAB.8.000685} {\bibfield  {journal} {\bibinfo
  {journal} {J. Opt. Soc. Am. B}\ }\textbf {\bibinfo {volume} {8}},\ \bibinfo
  {pages} {685} (\bibinfo {year} {1991}{\natexlab{a}})}\BibitemShut {NoStop}%
\bibitem [{\citenamefont {Gerry}\ and\ \citenamefont
  {Grobe}(1995)}]{Gerry1995}%
  \BibitemOpen
  \bibfield  {author} {\bibinfo {author} {\bibfnamefont {C.~C.}\ \bibnamefont
  {Gerry}}\ and\ \bibinfo {author} {\bibfnamefont {R.}~\bibnamefont {Grobe}},\
  }\bibfield  {title} {\bibinfo {title} {Two-mode intelligent {SU}(1,1)
  states},\ }\href {https://doi.org/10.1103/PhysRevA.51.4123} {\bibfield
  {journal} {\bibinfo  {journal} {Phys. Rev. A}\ }\textbf {\bibinfo {volume}
  {51}},\ \bibinfo {pages} {4123} (\bibinfo {year} {1995})}\BibitemShut
  {NoStop}%
\bibitem [{\citenamefont {Stoler}(1971)}]{stoler1971}%
  \BibitemOpen
  \bibfield  {author} {\bibinfo {author} {\bibfnamefont {D.}~\bibnamefont
  {Stoler}},\ }\bibfield  {title} {\bibinfo {title} {Generalized coherent
  states},\ }\href {https://doi.org/10.1103/PhysRevD.4.2309} {\bibfield
  {journal} {\bibinfo  {journal} {Phys. Rev. D}\ }\textbf {\bibinfo {volume}
  {4}},\ \bibinfo {pages} {2309} (\bibinfo {year} {1971})}\BibitemShut
  {NoStop}%
\bibitem [{\citenamefont {Yuen}(1976)}]{Yuen1976}%
  \BibitemOpen
  \bibfield  {author} {\bibinfo {author} {\bibfnamefont {H.~P.}\ \bibnamefont
  {Yuen}},\ }\bibfield  {title} {\bibinfo {title} {Two-photon coherent states
  of the radiation field},\ }\href {https://doi.org/10.1103/PhysRevA.13.2226}
  {\bibfield  {journal} {\bibinfo  {journal} {Phys. Rev. A}\ }\textbf {\bibinfo
  {volume} {13}},\ \bibinfo {pages} {2226} (\bibinfo {year}
  {1976})}\BibitemShut {NoStop}%
\bibitem [{\citenamefont {Gerry}(2001)}]{Gerry01}%
  \BibitemOpen
  \bibfield  {author} {\bibinfo {author} {\bibfnamefont {C.~C.}\ \bibnamefont
  {Gerry}},\ }\bibfield  {title} {\bibinfo {title} {Remarks on the use of group
  theory in quantum optics},\ }\href {https://doi.org/10.1364/OE.8.000076}
  {\bibfield  {journal} {\bibinfo  {journal} {Opt. Express}\ }\textbf {\bibinfo
  {volume} {8}},\ \bibinfo {pages} {76} (\bibinfo {year} {2001})}\BibitemShut
  {NoStop}%
\bibitem [{\citenamefont {Shaterzadeh-Yazdi}\ \emph {et~al.}(2008)\citenamefont
  {Shaterzadeh-Yazdi}, \citenamefont {Turner},\ and\ \citenamefont
  {Sanders}}]{yazdi2008}%
  \BibitemOpen
  \bibfield  {author} {\bibinfo {author} {\bibfnamefont {Z.}~\bibnamefont
  {Shaterzadeh-Yazdi}}, \bibinfo {author} {\bibfnamefont {P.~S.}\ \bibnamefont
  {Turner}},\ and\ \bibinfo {author} {\bibfnamefont {B.~C.}\ \bibnamefont
  {Sanders}},\ }\bibfield  {title} {\bibinfo {title} {{SU}(1,1) symmetry of
  multimode squeezed states},\ }\href
  {https://doi.org/10.1088/1751-8113/41/5/055309} {\bibfield  {journal}
  {\bibinfo  {journal} {J. Phys. A: Math. Theor.}\ }\textbf {\bibinfo {volume}
  {41}},\ \bibinfo {pages} {055309} (\bibinfo {year} {2008})}\BibitemShut
  {NoStop}%
\bibitem [{\citenamefont {Gilles}\ and\ \citenamefont
  {Knight}(1992)}]{luc1992}%
  \BibitemOpen
  \bibfield  {author} {\bibinfo {author} {\bibfnamefont {L.}~\bibnamefont
  {Gilles}}\ and\ \bibinfo {author} {\bibfnamefont {P.}~\bibnamefont
  {Knight}},\ }\bibfield  {title} {\bibinfo {title} {Non-classical properties
  of two-mode {SU}(1,1) coherent states},\ }\href
  {https://doi.org/10.1080/09500349214551471} {\bibfield  {journal} {\bibinfo
  {journal} {J. Mod. Opt.}\ }\textbf {\bibinfo {volume} {39}},\ \bibinfo
  {pages} {1411} (\bibinfo {year} {1992})}\BibitemShut {NoStop}%
\bibitem [{\citenamefont {Seyfarth}\ \emph {et~al.}(2020)\citenamefont
  {Seyfarth}, \citenamefont {Klimov}, \citenamefont {Guise}, \citenamefont
  {Leuchs},\ and\ \citenamefont {Sanchez-Soto}}]{Seyfarth2020}%
  \BibitemOpen
  \bibfield  {author} {\bibinfo {author} {\bibfnamefont {U.}~\bibnamefont
  {Seyfarth}}, \bibinfo {author} {\bibfnamefont {A.~B.}\ \bibnamefont
  {Klimov}}, \bibinfo {author} {\bibfnamefont {H.~d.}\ \bibnamefont {Guise}},
  \bibinfo {author} {\bibfnamefont {G.}~\bibnamefont {Leuchs}},\ and\ \bibinfo
  {author} {\bibfnamefont {L.~L.}\ \bibnamefont {Sanchez-Soto}},\ }\bibfield
  {title} {\bibinfo {title} {Wigner function for {SU}(1,1)},\ }\href
  {https://doi.org/10.22331/q-2020-09-07-317} {\bibfield  {journal} {\bibinfo
  {journal} {{Quantum}}\ }\textbf {\bibinfo {volume} {4}},\ \bibinfo {pages}
  {317} (\bibinfo {year} {2020})}\BibitemShut {NoStop}%
\bibitem [{\citenamefont {Klimov}\ \emph {et~al.}(2021)\citenamefont {Klimov},
  \citenamefont {Seyfarth}, \citenamefont {de~Guise},\ and\ \citenamefont
  {S{\'{a}}nchez-Soto}}]{Klimov2021}%
  \BibitemOpen
  \bibfield  {author} {\bibinfo {author} {\bibfnamefont {A.~B.}\ \bibnamefont
  {Klimov}}, \bibinfo {author} {\bibfnamefont {U.}~\bibnamefont {Seyfarth}},
  \bibinfo {author} {\bibfnamefont {H.}~\bibnamefont {de~Guise}},\ and\
  \bibinfo {author} {\bibfnamefont {L.~L.}\ \bibnamefont
  {S{\'{a}}nchez-Soto}},\ }\bibfield  {title} {\bibinfo {title} {{SU}(1, 1)
  covariant s-parametrized maps},\ }\href
  {https://doi.org/10.1088/1751-8121/abd7b4} {\bibfield  {journal} {\bibinfo
  {journal} {J. Phys. A: Math. Theor.}\ }\textbf {\bibinfo {volume} {54}},\
  \bibinfo {pages} {065301} (\bibinfo {year} {2021})}\BibitemShut {NoStop}%
\bibitem [{\citenamefont {Rundle}\ and\ \citenamefont
  {Everitt}(2021)}]{Russel2021}%
  \BibitemOpen
  \bibfield  {author} {\bibinfo {author} {\bibfnamefont {R.~P.}\ \bibnamefont
  {Rundle}}\ and\ \bibinfo {author} {\bibfnamefont {M.~J.}\ \bibnamefont
  {Everitt}},\ }\bibfield  {title} {\bibinfo {title} {Overview of the phase
  space formulation of quantum mechanics with application to quantum
  technologies},\ }\href
  {https://doi.org/https://doi.org/10.1002/qute.202100016} {\bibfield
  {journal} {\bibinfo  {journal} {Adv. Quantum Technol.}\ }\textbf {\bibinfo
  {volume} {4}},\ \bibinfo {pages} {2100016} (\bibinfo {year}
  {2021})}\BibitemShut {NoStop}%
\bibitem [{\citenamefont {Ban}(1994)}]{ban1994}%
  \BibitemOpen
  \bibfield  {author} {\bibinfo {author} {\bibfnamefont {M.}~\bibnamefont
  {Ban}},\ }\bibfield  {title} {\bibinfo {title} {Superpositions of the
  {SU}(1,1) coherent states},\ }\href
  {https://doi.org/https://doi.org/10.1016/0375-9601(94)90946-6} {\bibfield
  {journal} {\bibinfo  {journal} {Phys. Lett. A}\ }\textbf {\bibinfo {volume}
  {193}},\ \bibinfo {pages} {121} (\bibinfo {year} {1994})}\BibitemShut
  {NoStop}%
\bibitem [{\citenamefont {Gerry}\ and\ \citenamefont
  {Grobe}(1997)}]{Gerry1997}%
  \BibitemOpen
  \bibfield  {author} {\bibinfo {author} {\bibfnamefont {C.~C.}\ \bibnamefont
  {Gerry}}\ and\ \bibinfo {author} {\bibfnamefont {R.}~\bibnamefont {Grobe}},\
  }\bibfield  {title} {\bibinfo {title} {Two-mode {SU}(2) and {SU}(1,1)
  {S}chr\"{o}dinger cat states},\ }\href
  {https://doi.org/10.1080/09500349708232898} {\bibfield  {journal} {\bibinfo
  {journal} {J. Mod. Opt.}\ }\textbf {\bibinfo {volume} {44}},\ \bibinfo
  {pages} {41} (\bibinfo {year} {1997})}\BibitemShut {NoStop}%
\bibitem [{\citenamefont {Miry}\ and\ \citenamefont
  {Tavassoly}(2012)}]{Miry2012}%
  \BibitemOpen
  \bibfield  {author} {\bibinfo {author} {\bibfnamefont {S.~R.}\ \bibnamefont
  {Miry}}\ and\ \bibinfo {author} {\bibfnamefont {M.~K.}\ \bibnamefont
  {Tavassoly}},\ }\bibfield  {title} {\bibinfo {title} {Generation of a class
  of {SU}(1,1) coherent states of the {G}ilmore-{P}erelomov type and a class of
  {SU}(2) coherent states and their superposition},\ }\href
  {https://doi.org/10.1088/0031-8949/85/03/035404} {\bibfield  {journal}
  {\bibinfo  {journal} {Phys. Scr.}\ }\textbf {\bibinfo {volume} {85}},\
  \bibinfo {pages} {035404} (\bibinfo {year} {2012})}\BibitemShut {NoStop}%
\bibitem [{\citenamefont {Zheng}(2002)}]{zheng2002generation}%
  \BibitemOpen
  \bibfield  {author} {\bibinfo {author} {\bibfnamefont {S.-B.}\ \bibnamefont
  {Zheng}},\ }\bibfield  {title} {\bibinfo {title} {Generation of two-mode
  {SU}(2) and {SU}(1,1) cat states in a two-dimensional anisotropic ion trap},\
  }\href {https://doi.org/10.1023/A:1014564303100} {\bibfield  {journal}
  {\bibinfo  {journal} {Czech. J. Phys.}\ }\textbf {\bibinfo {volume} {52}},\
  \bibinfo {pages} {379} (\bibinfo {year} {2002})}\BibitemShut {NoStop}%
\bibitem [{\citenamefont {Sanders}(1992{\natexlab{a}})}]{San92}%
  \BibitemOpen
  \bibfield  {author} {\bibinfo {author} {\bibfnamefont {B.~C.}\ \bibnamefont
  {Sanders}},\ }\bibfield  {title} {\bibinfo {title} {Entangled coherent
  states},\ }\href {https://doi.org/10.1103/PhysRevA.45.6811} {\bibfield
  {journal} {\bibinfo  {journal} {Phys. Rev. A}\ }\textbf {\bibinfo {volume}
  {45}},\ \bibinfo {pages} {6811} (\bibinfo {year}
  {1992}{\natexlab{a}})}\BibitemShut {NoStop}%
\bibitem [{\citenamefont {Sanders}(1992{\natexlab{b}})}]{San92E}%
  \BibitemOpen
  \bibfield  {author} {\bibinfo {author} {\bibfnamefont {B.~C.}\ \bibnamefont
  {Sanders}},\ }\bibfield  {title} {\bibinfo {title} {Erratum: Entangled
  coherent states [{P}hys. {R}ev. {A} 45, 6811 (1992)]},\ }\href
  {https://doi.org/10.1103/PhysRevA.46.2966} {\bibfield  {journal} {\bibinfo
  {journal} {Phys. Rev. A}\ }\textbf {\bibinfo {volume} {46}},\ \bibinfo
  {pages} {2966} (\bibinfo {year} {1992}{\natexlab{b}})}\BibitemShut {NoStop}%
\bibitem [{\citenamefont {Sanders}(2012)}]{San12}%
  \BibitemOpen
  \bibfield  {author} {\bibinfo {author} {\bibfnamefont {B.~C.}\ \bibnamefont
  {Sanders}},\ }\bibfield  {title} {\bibinfo {title} {Review of entangled
  coherent states},\ }\href {https://doi.org/10.1088/1751-8113/45/24/244002}
  {\bibfield  {journal} {\bibinfo  {journal} {J. Phys. A: Math. Theor.}\
  }\textbf {\bibinfo {volume} {45}},\ \bibinfo {pages} {244002} (\bibinfo
  {year} {2012})}\BibitemShut {NoStop}%
\bibitem [{\citenamefont {Wang}\ \emph {et~al.}(2000)\citenamefont {Wang},
  \citenamefont {Sanders},\ and\ \citenamefont {Pan}}]{WSP00}%
  \BibitemOpen
  \bibfield  {author} {\bibinfo {author} {\bibfnamefont {X.}~\bibnamefont
  {Wang}}, \bibinfo {author} {\bibfnamefont {B.~C.}\ \bibnamefont {Sanders}},\
  and\ \bibinfo {author} {\bibfnamefont {S.-h.}\ \bibnamefont {Pan}},\
  }\bibfield  {title} {\bibinfo {title} {Entangled coherent states for systems
  with {SU}(2) and {SU}(1,1) symmetries},\ }\href
  {https://doi.org/10.1088/0305-4470/33/41/312} {\bibfield  {journal} {\bibinfo
   {journal} {J. Phys. A: Math. Gen.}\ }\textbf {\bibinfo {volume} {33}},\
  \bibinfo {pages} {7451} (\bibinfo {year} {2000})}\BibitemShut {NoStop}%
\bibitem [{\citenamefont {Schr\"{o}dinger}(1935)}]{Sch35}%
  \BibitemOpen
  \bibfield  {author} {\bibinfo {author} {\bibfnamefont {E.}~\bibnamefont
  {Schr\"{o}dinger}},\ }\bibfield  {title} {\bibinfo {title} {{Die
  gegenw{\"a}rtige {S}ituation in der {Q}uantenmechanik}},\ }\href
  {https://doi.org/10.1007/BF01491891} {\bibfield  {journal} {\bibinfo
  {journal} {Naturwissenschaften}\ }\textbf {\bibinfo {volume} {23}},\ \bibinfo
  {pages} {807} (\bibinfo {year} {1935})}\BibitemShut {NoStop}%
\bibitem [{\citenamefont {Navarrete-Benlloch}(2015)}]{CarlosNB15}%
  \BibitemOpen
  \bibfield  {author} {\bibinfo {author} {\bibfnamefont {C.}~\bibnamefont
  {Navarrete-Benlloch}},\ }\href@noop {} {\emph {\bibinfo {title} {An
  Introduction to the Formalism of Quantum Information with Continuous
  Variables}}}\ (\bibinfo  {publisher} {Morgan \& Claypool/IOP},\ \bibinfo
  {address} {Bristol},\ \bibinfo {year} {2015})\BibitemShut {NoStop}%
\bibitem [{\citenamefont {Audenaert}(2014)}]{Audenaert14}%
  \BibitemOpen
  \bibfield  {author} {\bibinfo {author} {\bibfnamefont {K.~M.~R.}\
  \bibnamefont {Audenaert}},\ }\bibfield  {title} {\bibinfo {title}
  {Comparisons between quantum state distinguishability measures},\ }\href
  {http://www.rintonpress.com/xxqic14/qic-14-12/0031-0038.pdf} {\bibfield
  {journal} {\bibinfo  {journal} {Quantum Inf. Comput.}\ }\textbf {\bibinfo
  {volume} {14}},\ \bibinfo {pages} {31} (\bibinfo {year} {2014})}\BibitemShut
  {NoStop}%
\bibitem [{\citenamefont {Sanders}(1989{\natexlab{a}})}]{San89}%
  \BibitemOpen
  \bibfield  {author} {\bibinfo {author} {\bibfnamefont {B.~C.}\ \bibnamefont
  {Sanders}},\ }\bibfield  {title} {\bibinfo {title} {Quantum dynamics of the
  nonlinear rotator and the effects of continual spin measurement},\ }\href
  {https://doi.org/10.1103/PhysRevA.40.2417} {\bibfield  {journal} {\bibinfo
  {journal} {Phys. Rev. A}\ }\textbf {\bibinfo {volume} {40}},\ \bibinfo
  {pages} {2417} (\bibinfo {year} {1989}{\natexlab{a}})}\BibitemShut {NoStop}%
\bibitem [{\citenamefont {Sanders}\ and\ \citenamefont {Gerry}(2014)}]{SG14}%
  \BibitemOpen
  \bibfield  {author} {\bibinfo {author} {\bibfnamefont {B.~C.}\ \bibnamefont
  {Sanders}}\ and\ \bibinfo {author} {\bibfnamefont {C.~C.}\ \bibnamefont
  {Gerry}},\ }\bibfield  {title} {\bibinfo {title} {Connection between the
  {NOON} state and a superposition of {SU}(2) coherent states},\ }\href
  {https://doi.org/10.1103/PhysRevA.90.045804} {\bibfield  {journal} {\bibinfo
  {journal} {Phys. Rev. A}\ }\textbf {\bibinfo {volume} {90}},\ \bibinfo
  {pages} {045804} (\bibinfo {year} {2014})}\BibitemShut {NoStop}%
\bibitem [{\citenamefont {Davis}\ \emph {et~al.}(2021)\citenamefont {Davis},
  \citenamefont {Kumari}, \citenamefont {Mann},\ and\ \citenamefont
  {Ghose}}]{davis2020}%
  \BibitemOpen
  \bibfield  {author} {\bibinfo {author} {\bibfnamefont {J.}~\bibnamefont
  {Davis}}, \bibinfo {author} {\bibfnamefont {M.}~\bibnamefont {Kumari}},
  \bibinfo {author} {\bibfnamefont {R.~B.}\ \bibnamefont {Mann}},\ and\
  \bibinfo {author} {\bibfnamefont {S.}~\bibnamefont {Ghose}},\ }\bibfield
  {title} {\bibinfo {title} {\uppercase{W}igner negativity in spin-$j$
  systems},\ }\href {https://doi.org/10.1103/PhysRevResearch.3.033134}
  {\bibfield  {journal} {\bibinfo  {journal} {Phys. Rev. Research}\ }\textbf
  {\bibinfo {volume} {3}},\ \bibinfo {pages} {033134} (\bibinfo {year}
  {2021})}\BibitemShut {NoStop}%
\bibitem [{\citenamefont {Huang}\ \emph {et~al.}(2015)\citenamefont {Huang},
  \citenamefont {Qin}, \citenamefont {Zhong}, \citenamefont {Ke},\ and\
  \citenamefont {Lee}}]{Huang15}%
  \BibitemOpen
  \bibfield  {author} {\bibinfo {author} {\bibfnamefont {J.}~\bibnamefont
  {Huang}}, \bibinfo {author} {\bibfnamefont {X.}~\bibnamefont {Qin}}, \bibinfo
  {author} {\bibfnamefont {H.}~\bibnamefont {Zhong}}, \bibinfo {author}
  {\bibfnamefont {Y.}~\bibnamefont {Ke}},\ and\ \bibinfo {author}
  {\bibfnamefont {C.}~\bibnamefont {Lee}},\ }\bibfield  {title} {\bibinfo
  {title} {Quantum metrology with spin cat states under dissipation},\ }\href
  {https://doi.org/10.1038/srep17894} {\bibfield  {journal} {\bibinfo
  {journal} {Sci. Rep.}\ }\textbf {\bibinfo {volume} {5}},\ \bibinfo {pages}
  {1} (\bibinfo {year} {2015})}\BibitemShut {NoStop}%
\bibitem [{\citenamefont {Huang}\ \emph {et~al.}(2018)\citenamefont {Huang},
  \citenamefont {Zhuang}, \citenamefont {Lu}, \citenamefont {Ke},\ and\
  \citenamefont {Lee}}]{Huang18}%
  \BibitemOpen
  \bibfield  {author} {\bibinfo {author} {\bibfnamefont {J.}~\bibnamefont
  {Huang}}, \bibinfo {author} {\bibfnamefont {M.}~\bibnamefont {Zhuang}},
  \bibinfo {author} {\bibfnamefont {B.}~\bibnamefont {Lu}}, \bibinfo {author}
  {\bibfnamefont {Y.}~\bibnamefont {Ke}},\ and\ \bibinfo {author}
  {\bibfnamefont {C.}~\bibnamefont {Lee}},\ }\bibfield  {title} {\bibinfo
  {title} {Achieving \uppercase{h}eisenberg-limited metrology with spin cat
  states via interaction-based readout},\ }\href
  {https://doi.org/10.1103/PhysRevA.98.012129} {\bibfield  {journal} {\bibinfo
  {journal} {Phys. Rev. A}\ }\textbf {\bibinfo {volume} {98}},\ \bibinfo
  {pages} {012129} (\bibinfo {year} {2018})}\BibitemShut {NoStop}%
\bibitem [{\citenamefont {Maleki}\ and\ \citenamefont
  {Zheltikov}(2020)}]{Maleki}%
  \BibitemOpen
  \bibfield  {author} {\bibinfo {author} {\bibfnamefont {Y.}~\bibnamefont
  {Maleki}}\ and\ \bibinfo {author} {\bibfnamefont {A.~M.}\ \bibnamefont
  {Zheltikov}},\ }\bibfield  {title} {\bibinfo {title} {Spin cat-state family
  for \uppercase{h}eisenberg-limit metrology},\ }\href
  {https://doi.org/10.1364/JOSAB.374221} {\bibfield  {journal} {\bibinfo
  {journal} {J. Opt. Soc. Am. B}\ }\textbf {\bibinfo {volume} {37}},\ \bibinfo
  {pages} {1021} (\bibinfo {year} {2020})}\BibitemShut {NoStop}%
\bibitem [{\citenamefont {Varilly}\ and\ \citenamefont
  {Gracia-Bond}(1989)}]{Varilly89}%
  \BibitemOpen
  \bibfield  {author} {\bibinfo {author} {\bibfnamefont {J.~C.}\ \bibnamefont
  {Varilly}}\ and\ \bibinfo {author} {\bibfnamefont {J.~M.}\ \bibnamefont
  {Gracia-Bond}},\ }\bibfield  {title} {\bibinfo {title} {The {M}oyal
  representation for spin},\ }\href
  {https://doi.org/https://doi.org/10.1016/0003-4916(89)90262-5} {\bibfield
  {journal} {\bibinfo  {journal} {Ann. Phys. (NY)}\ }\textbf {\bibinfo {volume}
  {190}},\ \bibinfo {pages} {107} (\bibinfo {year} {1989})}\BibitemShut
  {NoStop}%
\bibitem [{\citenamefont {Heiss}\ and\ \citenamefont
  {Weigert}(2000)}]{Heiss00}%
  \BibitemOpen
  \bibfield  {author} {\bibinfo {author} {\bibfnamefont {S.}~\bibnamefont
  {Heiss}}\ and\ \bibinfo {author} {\bibfnamefont {S.}~\bibnamefont
  {Weigert}},\ }\bibfield  {title} {\bibinfo {title} {Discrete
  \uppercase{m}oyal-type representations for a spin},\ }\href
  {https://doi.org/10.1103/PhysRevA.63.012105} {\bibfield  {journal} {\bibinfo
  {journal} {Phys. Rev. A}\ }\textbf {\bibinfo {volume} {63}},\ \bibinfo
  {pages} {012105} (\bibinfo {year} {2000})}\BibitemShut {NoStop}%
\bibitem [{\citenamefont {Klimov}\ \emph {et~al.}(2017)\citenamefont {Klimov},
  \citenamefont {Romero},\ and\ \citenamefont {de~Guise}}]{Klimov17}%
  \BibitemOpen
  \bibfield  {author} {\bibinfo {author} {\bibfnamefont {A.~B.}\ \bibnamefont
  {Klimov}}, \bibinfo {author} {\bibfnamefont {J.~L.}\ \bibnamefont {Romero}},\
  and\ \bibinfo {author} {\bibfnamefont {H.}~\bibnamefont {de~Guise}},\
  }\bibfield  {title} {\bibinfo {title} {Generalized {SU}(2) covariant
  \uppercase{W}igner functions and some of their applications},\ }\href
  {https://doi.org/10.1088/1751-8121/50/32/323001} {\bibfield  {journal}
  {\bibinfo  {journal} {J. Phys. A: Math. Theor.}\ }\textbf {\bibinfo {volume}
  {50}},\ \bibinfo {pages} {323001} (\bibinfo {year} {2017})}\BibitemShut
  {NoStop}%
\bibitem [{\citenamefont {Koczor}\ \emph {et~al.}(2020)\citenamefont {Koczor},
  \citenamefont {Zeier},\ and\ \citenamefont {Glaser}}]{Glaser20}%
  \BibitemOpen
  \bibfield  {author} {\bibinfo {author} {\bibfnamefont {B.}~\bibnamefont
  {Koczor}}, \bibinfo {author} {\bibfnamefont {R.}~\bibnamefont {Zeier}},\ and\
  \bibinfo {author} {\bibfnamefont {S.~J.}\ \bibnamefont {Glaser}},\ }\bibfield
   {title} {\bibinfo {title} {Continuous phase-space representations for
  finite-dimensional quantum states and their tomography},\ }\href
  {https://doi.org/10.1103/PhysRevA.101.022318} {\bibfield  {journal} {\bibinfo
   {journal} {Phys. Rev. A}\ }\textbf {\bibinfo {volume} {101}},\ \bibinfo
  {pages} {022318} (\bibinfo {year} {2020})}\BibitemShut {NoStop}%
\bibitem [{\citenamefont {Loudon}\ and\ \citenamefont
  {Knight}(1987)}]{Knight1987}%
  \BibitemOpen
  \bibfield  {author} {\bibinfo {author} {\bibfnamefont {R.}~\bibnamefont
  {Loudon}}\ and\ \bibinfo {author} {\bibfnamefont {P.}~\bibnamefont
  {Knight}},\ }\bibfield  {title} {\bibinfo {title} {Squeezed light},\ }\href
  {https://doi.org/10.1080/09500348714550721} {\bibfield  {journal} {\bibinfo
  {journal} {J. Mod. Opt.}\ }\textbf {\bibinfo {volume} {34}},\ \bibinfo
  {pages} {709} (\bibinfo {year} {1987})}\BibitemShut {NoStop}%
\bibitem [{\citenamefont {Giovannetti}\ \emph {et~al.}(2004)\citenamefont
  {Giovannetti}, \citenamefont {Lloyd},\ and\ \citenamefont
  {Maccone}}]{Maccone2004}%
  \BibitemOpen
  \bibfield  {author} {\bibinfo {author} {\bibfnamefont {V.}~\bibnamefont
  {Giovannetti}}, \bibinfo {author} {\bibfnamefont {S.}~\bibnamefont {Lloyd}},\
  and\ \bibinfo {author} {\bibfnamefont {L.}~\bibnamefont {Maccone}},\
  }\bibfield  {title} {\bibinfo {title} {Quantum-enhanced measurements: Beating
  the standard quantum limit},\ }\href
  {https://doi.org/10.1126/science.1104149} {\bibfield  {journal} {\bibinfo
  {journal} {Science}\ }\textbf {\bibinfo {volume} {306}},\ \bibinfo {pages}
  {1330} (\bibinfo {year} {2004})}\BibitemShut {NoStop}%
\bibitem [{\citenamefont {Mufti}\ \emph {et~al.}(1993)\citenamefont {Mufti},
  \citenamefont {Schmitt},\ and\ \citenamefont {Sargent}}]{mufti1993}%
  \BibitemOpen
  \bibfield  {author} {\bibinfo {author} {\bibfnamefont {A.}~\bibnamefont
  {Mufti}}, \bibinfo {author} {\bibfnamefont {H.~A.}\ \bibnamefont {Schmitt}},\
  and\ \bibinfo {author} {\bibfnamefont {M.}~\bibnamefont {Sargent}},\
  }\bibfield  {title} {\bibinfo {title} {Finite‐dimensional matrix
  representations as calculational tools in quantum optics},\ }\href
  {https://doi.org/10.1119/1.17149} {\bibfield  {journal} {\bibinfo  {journal}
  {Am. J. Phys.}\ }\textbf {\bibinfo {volume} {61}},\ \bibinfo {pages} {729}
  (\bibinfo {year} {1993})}\BibitemShut {NoStop}%
\bibitem [{\citenamefont {Fujii}(2001)}]{fujii2001introduction}%
  \BibitemOpen
  \bibfield  {author} {\bibinfo {author} {\bibfnamefont {K.}~\bibnamefont
  {Fujii}},\ }\bibfield  {title} {\bibinfo {title} {Introduction to coherent
  states and quantum information theory},\ }\href@noop {} {\bibfield  {journal}
  {\bibinfo  {journal} {arXiv e-prints}\ ,\ \bibinfo {pages} {quant}} (\bibinfo
  {year} {2001})}\BibitemShut {NoStop}%
\bibitem [{\citenamefont {Hach}\ \emph {et~al.}(2018)\citenamefont {Hach},
  \citenamefont {Birrittella}, \citenamefont {Alsing},\ and\ \citenamefont
  {Gerry}}]{edwin2018}%
  \BibitemOpen
  \bibfield  {author} {\bibinfo {author} {\bibfnamefont {E.~E.}\ \bibnamefont
  {Hach}}, \bibinfo {author} {\bibfnamefont {R.}~\bibnamefont {Birrittella}},
  \bibinfo {author} {\bibfnamefont {P.~M.}\ \bibnamefont {Alsing}},\ and\
  \bibinfo {author} {\bibfnamefont {C.~C.}\ \bibnamefont {Gerry}},\ }\bibfield
  {title} {\bibinfo {title} {{SU}(1,1) parity and strong violations of a {B}ell
  inequality by entangled {B}arut; {G}irardello coherent states},\ }\href
  {https://doi.org/10.1364/JOSAB.35.002433} {\bibfield  {journal} {\bibinfo
  {journal} {J. Opt. Soc. Am. B}\ }\textbf {\bibinfo {volume} {35}},\ \bibinfo
  {pages} {2433} (\bibinfo {year} {2018})}\BibitemShut {NoStop}%
\bibitem [{\citenamefont {{B}ing Tang}\ \emph {et~al.}(2015)\citenamefont
  {{B}ing Tang}, \citenamefont {Gao}, \citenamefont {{Y}ao-{X}iong Wang},
  \citenamefont {{G}uang Wu},\ and\ \citenamefont {Shuang}}]{TANG201586}%
  \BibitemOpen
  \bibfield  {author} {\bibinfo {author} {\bibfnamefont {X.}~\bibnamefont
  {{B}ing Tang}}, \bibinfo {author} {\bibfnamefont {F.}~\bibnamefont {Gao}},
  \bibinfo {author} {\bibnamefont {{Y}ao-{X}iong Wang}}, \bibinfo {author}
  {\bibfnamefont {J.}~\bibnamefont {{G}uang Wu}},\ and\ \bibinfo {author}
  {\bibfnamefont {F.}~\bibnamefont {Shuang}},\ }\bibfield  {title} {\bibinfo
  {title} {Non-{G}aussian features from excited squeezed vacuum state},\ }\href
  {https://doi.org/https://doi.org/10.1016/j.optcom.2015.01.053} {\bibfield
  {journal} {\bibinfo  {journal} {Opt. Commun.}\ }\textbf {\bibinfo {volume}
  {345}},\ \bibinfo {pages} {86} (\bibinfo {year} {2015})}\BibitemShut
  {NoStop}%
\bibitem [{\citenamefont {Sanders}(1989{\natexlab{b}})}]{barry1989}%
  \BibitemOpen
  \bibfield  {author} {\bibinfo {author} {\bibfnamefont {B.~C.}\ \bibnamefont
  {Sanders}},\ }\bibfield  {title} {\bibinfo {title} {Superposition of two
  squeezed vacuum states and interference effects},\ }\href
  {https://doi.org/10.1103/PhysRevA.39.4284} {\bibfield  {journal} {\bibinfo
  {journal} {Phys. Rev. A}\ }\textbf {\bibinfo {volume} {39}},\ \bibinfo
  {pages} {4284} (\bibinfo {year} {1989}{\natexlab{b}})}\BibitemShut {NoStop}%
\bibitem [{\citenamefont {de~Freitas}\ and\ \citenamefont
  {Dodonov}(2021)}]{quantumrep2021}%
  \BibitemOpen
  \bibfield  {author} {\bibinfo {author} {\bibfnamefont {M.~C.}\ \bibnamefont
  {de~Freitas}}\ and\ \bibinfo {author} {\bibfnamefont {V.~V.}\ \bibnamefont
  {Dodonov}},\ }\bibfield  {title} {\bibinfo {title} {Non-{G}aussianity of
  four-photon superpositions of {F}ock states},\ }\href
  {https://doi.org/10.3390/quantum3030022} {\bibfield  {journal} {\bibinfo
  {journal} {Quantum Reports}\ }\textbf {\bibinfo {volume} {3}},\ \bibinfo
  {pages} {350} (\bibinfo {year} {2021})}\BibitemShut {NoStop}%
\bibitem [{\citenamefont {Happ}\ \emph {et~al.}(2018)\citenamefont {Happ},
  \citenamefont {Efremov}, \citenamefont {Nha},\ and\ \citenamefont
  {Schleich}}]{Happ2018}%
  \BibitemOpen
  \bibfield  {author} {\bibinfo {author} {\bibfnamefont {L.}~\bibnamefont
  {Happ}}, \bibinfo {author} {\bibfnamefont {M.~A.}\ \bibnamefont {Efremov}},
  \bibinfo {author} {\bibfnamefont {H.}~\bibnamefont {Nha}},\ and\ \bibinfo
  {author} {\bibfnamefont {W.~P.}\ \bibnamefont {Schleich}},\ }\bibfield
  {title} {\bibinfo {title} {Sufficient condition for a quantum state to be
  genuinely quantum non-{G}aussian},\ }\href
  {https://doi.org/10.1088/1367-2630/aaac25} {\bibfield  {journal} {\bibinfo
  {journal} {New J. Phys.}\ }\textbf {\bibinfo {volume} {20}},\ \bibinfo
  {pages} {023046} (\bibinfo {year} {2018})}\BibitemShut {NoStop}%
\bibitem [{\citenamefont {Quesne}(2001)}]{quesene}%
  \BibitemOpen
  \bibfield  {author} {\bibinfo {author} {\bibfnamefont {C.}~\bibnamefont
  {Quesne}},\ }\bibfield  {title} {\bibinfo {title} {Completeness of
  photon-added squeezed vacuum and one-photon states and of photon-added
  coherent states on a circle},\ }\href
  {https://doi.org/https://doi.org/10.1016/S0375-9601(01)00554-0} {\bibfield
  {journal} {\bibinfo  {journal} {Phys. Lett. A}\ }\textbf {\bibinfo {volume}
  {288}},\ \bibinfo {pages} {241} (\bibinfo {year} {2001})}\BibitemShut
  {NoStop}%
\bibitem [{\citenamefont {Cardoso}\ \emph {et~al.}(2021)\citenamefont
  {Cardoso}, \citenamefont {Rossatto}, \citenamefont {Fernandes}, \citenamefont
  {Higgins},\ and\ \citenamefont {Villas-Boas}}]{Cardoso2021}%
  \BibitemOpen
  \bibfield  {author} {\bibinfo {author} {\bibfnamefont {F.~R.}\ \bibnamefont
  {Cardoso}}, \bibinfo {author} {\bibfnamefont {D.~Z.}\ \bibnamefont
  {Rossatto}}, \bibinfo {author} {\bibfnamefont {G.~P. L.~M.}\ \bibnamefont
  {Fernandes}}, \bibinfo {author} {\bibfnamefont {G.}~\bibnamefont {Higgins}},\
  and\ \bibinfo {author} {\bibfnamefont {C.~J.}\ \bibnamefont {Villas-Boas}},\
  }\bibfield  {title} {\bibinfo {title} {Superposition of two-mode squeezed
  states for quantum information processing and quantum sensing},\ }\href
  {https://doi.org/10.1103/PhysRevA.103.062405} {\bibfield  {journal} {\bibinfo
   {journal} {Phys. Rev. A}\ }\textbf {\bibinfo {volume} {103}},\ \bibinfo
  {pages} {062405} (\bibinfo {year} {2021})}\BibitemShut {NoStop}%
\bibitem [{\citenamefont {Gerry}(1991{\natexlab{b}})}]{Gerry91b}%
  \BibitemOpen
  \bibfield  {author} {\bibinfo {author} {\bibfnamefont {C.~C.}\ \bibnamefont
  {Gerry}},\ }\bibfield  {title} {\bibinfo {title} {Correlated two-mode
  {SU}(1,1) coherent states: nonclassical properties: errata},\ }\href
  {https://doi.org/10.1364/JOSAB.8.001999} {\bibfield  {journal} {\bibinfo
  {journal} {J. Opt. Soc. Am. B}\ }\textbf {\bibinfo {volume} {8}},\ \bibinfo
  {pages} {1999} (\bibinfo {year} {1991}{\natexlab{b}})}\BibitemShut {NoStop}%
\bibitem [{\citenamefont {Basit}\ \emph {et~al.}(2021)\citenamefont {Basit},
  \citenamefont {Ali}, \citenamefont {Badshah}, \citenamefont {Yang},\ and\
  \citenamefont {Ge}}]{Basit2021}%
  \BibitemOpen
  \bibfield  {author} {\bibinfo {author} {\bibfnamefont {A.}~\bibnamefont
  {Basit}}, \bibinfo {author} {\bibfnamefont {H.}~\bibnamefont {Ali}}, \bibinfo
  {author} {\bibfnamefont {F.}~\bibnamefont {Badshah}}, \bibinfo {author}
  {\bibfnamefont {X.-F.}\ \bibnamefont {Yang}},\ and\ \bibinfo {author}
  {\bibfnamefont {G.-Q.}\ \bibnamefont {Ge}},\ }\bibfield  {title} {\bibinfo
  {title} {Controlling sudden transition between classical and quantum
  decoherence via squeezing phase of the baths},\ }\href
  {https://doi.org/10.1088/1612-202x/abfa8c} {\bibfield  {journal} {\bibinfo
  {journal} {Laser Phys. Lett.}\ }\textbf {\bibinfo {volume} {18}},\ \bibinfo
  {pages} {065202} (\bibinfo {year} {2021})}\BibitemShut {NoStop}%
\bibitem [{\citenamefont {Bartlett}\ \emph {et~al.}(2001)\citenamefont
  {Bartlett}, \citenamefont {Rice}, \citenamefont {Sanders}, \citenamefont
  {Daboul},\ and\ \citenamefont {de~Guise}}]{PhysRevA.63.042310}%
  \BibitemOpen
  \bibfield  {author} {\bibinfo {author} {\bibfnamefont {S.~D.}\ \bibnamefont
  {Bartlett}}, \bibinfo {author} {\bibfnamefont {D.~A.}\ \bibnamefont {Rice}},
  \bibinfo {author} {\bibfnamefont {B.~C.}\ \bibnamefont {Sanders}}, \bibinfo
  {author} {\bibfnamefont {J.}~\bibnamefont {Daboul}},\ and\ \bibinfo {author}
  {\bibfnamefont {H.}~\bibnamefont {de~Guise}},\ }\bibfield  {title} {\bibinfo
  {title} {Unitary transformations for testing {B}ell inequalities},\ }\href
  {https://doi.org/10.1103/PhysRevA.63.042310} {\bibfield  {journal} {\bibinfo
  {journal} {Phys. Rev. A}\ }\textbf {\bibinfo {volume} {63}},\ \bibinfo
  {pages} {042310} (\bibinfo {year} {2001})}\BibitemShut {NoStop}%
\end{thebibliography}%
\end{document}